%% file: sn2023ixf_nebular_UV.tex
\documentclass[twocolumn]{aastex631}
\usepackage{CJK}
\usepackage[utf8]{inputenc}
\usepackage{multirow}
\usepackage{soul}
\newcommand\N[1]{{\color{blue} \bf #1}}

\defcitealias{astropy_collaboration_astropy_2013}{Astropy Collaboration 2013}
\defcitealias{astropy_collaboration_astropy_2018}{2018}
\defcitealias{astropy_collaboration_astropy_2022}{2022}

\input{new-commands}

\shorttitle{Late-time UV Spectra of SN~2023ixf, SN~2024ggi}
\shortauthors{Bostroem et al. 2025}

\begin{document}

\title{Late-time Hubble Space Telescope Ultraviolet Spectra of SN~2023ixf and SN~2024ggi Show Ongoing Interaction with Circumstellar Material}

\input{affiliations}
\input{authors}

\begin{abstract}
We present far- and near-ultraviolet (UV) spectra of the Type II supernovae (SNe) SN~2023ixf from days 199 to 722 and SN~2024ggi at days 41 and 232. 
Both supernovae show broad, blueshifted, and asymmetric UV emission lines with an initial maximum velocity of $\sim$9000 \kms\ and narrow unresolved emission in \CIVlambda.
We compare the optical and UV emission-line profiles, showing that they evolve from two distinct velocity profiles to a single profile tracing the UV emission.
We interpret this as shock power from interaction with circumstellar material coming to dominate over the radioactive-decay power from the inner ejecta.
Comparing our observations to radiative transfer models with injected shock power, we find SN~2024ggi is best matched by $P_{\mathrm{shock, abs}}=1\times10^{41}$ \pwrunit\ at day 40, SN~2023ixf at day 300 and SN~2024ggi at day 200 are best matched by $P_{\mathrm{shock,abs}}=1\times10^{40}$ \pwrunit, and SN~2023ixf at day 600 is best matched by $P_{\mathrm{shock,abs}}=5\times10^{39}$ \pwrunit. 
From these models, we find the mass-loss rate of both supernovae increased just before explosion. For SN~2023ixf our mass-loss rates go from  $4\times10^{-5}$ \masslossunit\ at 600 yr before explosion to $2\times10^{-2}$ \masslossunit\ at 15 yr prior to explosion. 
For SN~2024ggi, we find a mass-loss rate of $9\times10^{-5}$\masslossunit\ at 150 yr before explosion and $1\times10^{-3}$\masslossunit\ at 30 yr before explosion.

\end{abstract}

\section{Introduction} \label{sec:intro}
Core-collapse supernovae (SNe) are the end of life explosions of massive stars ($M\gtrsim8$\,\msun).
Over the course of their lifetimes, massive stars lose mass through stellar winds and/or binary interaction \citep[see][]{2014SmithARAA}.
When a star explodes with at least some of its hydrogen envelope still intact, we refer to the resulting supernova as Type II.
Type II supernovae whose light curve exhibits a rapid rise and a relatively flat plateau for $\sim$80--100 days, a fall from this plateau onto a linearly declining (in mag day$^{-1}$) radioactive decay powered tail are called Type IIP. 
In Type IIL supernovae, by contrast, the main light curve peak linearly declines (in mag day$^{-1}$) rather than staying at a constant brightness 
\citep{1979Barbon}. 
Type IIP/IIL supernovae appear to be two ends of a continuum of slopes \citep{2014Anderson}, and we do not distinguish between the subtypes in this paper.

All pre-explosion direct detections of Type IIP/L progenitors have been consistent with red supergiants  \citep[RSGs; ][]{2009Smartt, 2015Smartt}. 
While prescriptions for quiescent mass loss in RSGs exist, these disagree by up to three orders of magnitude, with mass-loss rates ranging from $\dot{M} =10^{-10} - 10^{-7}$ \masslossunit\ at the low-mass end to $\dot{M}=10^{-6}-10^{-4}$\masslossunit\ at the high-mass end \citep{2012Ekstrom,2020Beasor}.
Additionally, recent constraints on mass loss just before explosion from supernova light-curve modeling and narrow emission lines in early-time spectra   \citep{2016Khazov,morozova2018, 2023Subrayan,2023Bruch,2024Jacobson-Galan,2025Jacobson-Galan} identify even larger mass-loss rates.
Some physical mechanisms have been proposed such as pulsationally driven winds, nuclear burning instabilities, gravity-wave-driven mass loss, and super-Eddington mass loss \citep{2010Yoon, 1997Heger,2006Smith,2011Arnett,2012Quataert,2012Ekstrom,2014Smith, 2014Shiode,2017Fuller, 2021Wu,2022Wu}.
However, none has been definitively identified that accounts for both the amount of mass loss and the timing of the mass loss in Type IIP/L supernovae.

As a star loses mass, the material lost most recently may still be distributed around the star as circumstellar matter (CSM). 
Thus, one way to study mass loss in RSGs is to study the CSM surrounding the star.
After explosion, as a supernova expands, its ejecta encounter CSM that is progressively farther away from the star, with different epochs of observation probing different epochs of mass loss \citep[see][for a review]{2017Smith}.

Some RSG supernova progenitors appear to have dense, confined material near the stellar surface which manifests as narrow, high-ionization spectroscopic emission (sometimes with broad wings) in the early-time supernova spectra that fade within days to weeks after explosion \citep{ 2016Khazov, 2000Leonard, 1979Barbon, 2014Gal-Yam, 2015Smith,  2016Terreran, Yaron17, 2017Dessart,2018Hosseinzadeh, 2019Boian, 2021Tartaglia,2022Terreran,2022Jacobson-Galan, 2023Bruch, 2023Zimmerman,2023Vasylyev2, 2023Smith, 2023Hiramatsu,2023Jacobson-Galan, 2023Bostroem2,2024Jacobson-Galan2, 2024Andrews,2024Shrestha,2024Jacobson-Galan, 2025Dickinson, 2025Zheng, 2025Andrews}. 
These lines (sometimes referred to as flash ionization features) are powered by the loss in kinetic energy as the supernova ejecta run into dense CSM and thus probe RSGs with a period of high mass loss just prior to explosion.
Because of the transient nature of these lines, the characteristics of the populations that show them are still being explored. 
It is not clear if the subclass of Type IIP/L supernovae with these transient, narrow features originates from a particular region of parameter space (e.g., higher mass).
It is also unknown if it is a distinct class at all or part of a continuum of properties, including the mass-loss history of the progenitor, beyond the few years just before explosion probed by the narrow lines \citep{2025Jacobson-Galan, 2024Jacobson-Galan, 2023Bruch, 2016Khazov}.  

The UV is particularly sensitive to CSM interaction and can be used to map the mass-loss history of RSGs using early and late-time spectra. 
Although are rare, early UV spectra of SN~2021yja \citep{2022Vasylyev}, SN~2022wsp \citep{2023Vasylyev}, SN~2022acko \citep{2023Bostroem1}, SN~1999em \citep{2000Baron}, and SN~2005cs (Rowe et al. in prep) show a broad diversity which could reflect diverse RSG mass-loss histories. 
Late-time UV spectra look back even father into RSG mass-loss histories. 
When no CSM is present, this late phase is powered by the radioactive decay of $^{56}$Co in the inner supernova ejecta which emits in the optical, resulting in virtually no UV radiation. 
However, when the supernova ejecta encounter dense CSM, forward and reverse shock waves form and propagate out into the CSM and back into the supernova ejecta, respectively.
At the contact discontinuity between the two shocks, a cool, dense shell (CDS) can form.
Both the forward shock and the reverse shock emit X-rays which can be absorbed by the CDS and reprocessed into UV and optical light \citep{1994Chevalier}.
At late times, the supernova ejecta spectral energy distribution (SED) shifts from the UV into the optical. 
Modeling of CSM interaction at this phase suggests that, if CSM is present, it will dominate the UV flux \citep{2023Dessart,2022Dessart}.

SN~2023ixf and SN~2024ggi were two very nearby supernovae that were first detected within a year of each other.
SN~2023ixf was discovered on 2023-05-19 17:27:15 UTC \citep[all times in this paper are given in UTC; ][]{itagaki_discovery_2023}. 
In M101 at a distance of 6.9 Mpc \citep{2022Riess}, it was the closest Type IIP/L supernova in at least a decade.
It was rapidly classified and the first spectrum showed narrow high-ionization features indicating CSM interaction \citep{perley_classification_2023} which faded within the first 7 days \citep{2023Bostroem2, 2023Jacobson-Galan, 2023Hiramatsu, 2023Smith, 2023Vasylyev2, 2023Zimmerman, 2025Zheng, 2025Dickinson}. 
SN~2023ixf is one of the best-studied Type IIP/L supernovae,  with observations spanning the entire electromagnetic spectrum as well as neutrino and gravitational-wave searches \citep[e.g.,][]{2024Ravensburg, 2024Marti-Devesa,2023Guetta, 2023Berger,2025Nayana, 2024Sarmah,2025Iwata, 2025Kimura,2025Abac, 2025Szczepanczyk, 2025Cosentino}. 
Of particular note is the unprecedented UV time series, starting with the earliest near-UV (NUV) observations of an Type IIP/L supernova at day $\sim$3. 
These observations showed narrow P Cygni lines which give way to broad P Cygni profiles by day 8; a broad, boxy \MgII\ emission line develops around day 20 and persists until at least day 66 \citep{2023Teja, 2024Bostroem,2023Zimmerman}. 
We adopt an explosion epoch of 2023-05-18 18:00:00 ($\pm0.1$\,d) \citep{2023Hosseinzadeh}, a Milky Way extinction $E(B-V)=0.0076\pm0.0002$\,mag \citep{2011Schlafly}, and a host-galaxy extinction $E(B-V)=0.031\pm0.01$\,mag \citep{2023Smith}.

On 2024-04-11 03:21:36, SN~2024ggi was discovered in NGC 3621 \citep[$7.24\pm0.2$ Mpc;][]{2006Saha, 2024Srivastav}. 
SN~2024ggi showed narrow, high-ionization features similar to those of SN~2023ixf but which disappeared faster \citep[by day 3.6;][]{2024Jacobson-Galan2,2024Shrestha}. 
SN~2024ggi thus provides a rare opportunity to compare two very nearby supernovae having dense, confined CSM. 
From \citet{2024Shrestha}, we adopt an explosion epoch of 2024-04-10 19:12:00 for SN~2024ggi, a Milky Way extinction $E(B-V)=0.12\pm0.028$\,mag, and a host-galaxy extinction $E(B-V)=0.034\pm0.020$\,mag. 

Here we present UV spectra of SN~2023ixf from days $\sim$200 to 700 and SN~2024ggi at days $\sim$40 and 200 observed by the Space Telescope Imaging Spectrograph (STIS) and the Cosmic Origins Spectrograph (COS) on the Hubble Space Telescope (HST). 
These represent the first HST UV spectra of SN~2024ggi and the first far-UV (FUV) spectra of SN~2023ixf. 
In \autoref{sec:obs} we describe our observations and data reduction.
We identify lines and analyze the line profiles in \autoref{sec:analysis} and compare our observations to radiative transfer models in \autoref{sec:models}. 
Finally, in \autoref{sec:discussion} we derive a mass-loss history from our best-matched models and compare the spectra of SN~2023ixf and SN~2024ggi to all other Type II supernovae that have been observed by HST in the UV at comparable epochs. 
We close with a summary of our results in \autoref{sec:summary}.

\section{Observations}\label{sec:obs}
\subsection{UV Spectra}
UV observations of SN~2023ixf were performed with STIS and COS onboard HST from $\sim$66 to 720\,d as part of proposals 17313 (PI K. A. Bostroem), 17497 (PI S. Valenti), and 17772 (PI K. A. Bostroem).
A complete list of successful observations is given in \autoref{tab:observation}.
At day 66, the STIS CCD was used to observe the NUV through near-infrared (NIR) with the G230LB, G430L, and G750L gratings. 
The reduction of these observations is described in detail by \citet{2024Bostroem}.
Briefly, all spectra were reduced with \texttt{calstis}. 
The G430L data were automatically reduced by the Mikulski Archive for Space Telescopes (MAST).
The G750L spectrum was reduced with default parameters including the application of a fringe-flat.
The location of the trace in the G230LB spectrum was identified manually before the automated extraction was applied. 

FUV (STIS/G140L) observations of SN~2023ixf were obtained between days 183 and 723 at four epochs.
NUV (STIS/G230L and COS/G230L) observations of SN~2023ixf were obtained between days 66 and 723 at five epochs.
At phases $>200$\,d, the UV spectra are evolving slowly, so we combine FUV and NUV observations at similar epochs into single observations with average epochs of 199\,d, 311\,d, 620\,d, and 722\,d using the Hubble Advanced Spectral Product coadd script \citep{2024Debes}.
All data were calibrated with \texttt{calstis} and \texttt{calcos}, and we used the one-dimensional (1D) calibrated spectra downloaded from the MAST unless otherwise noted. 
We cut out regions of the spectrum strongly affected by airglow (\Lya\ and \OIlambda) or interstellar matter (ISM) absorption (\MgIIlambda).
We identify the regions affected by airglow using the \texttt{background} array in the 1D extracted files, which shows the flux of the region above and below the spectrum extraction box. 
The strength of the airglow lines varies with time, so the region affected is different at each epoch. 
In SN~2023ixf and SN~2024ggi at these phases, the ISM absorption is easily identifiable as narrow absorption on top of a broad supernova feature. 

Whenever possible, we observed with STIS, as its narrow slit reduces contamination that could be possible within the $2.5''$ circular aperture of COS without significant loss of sensitivity (in low-resolution modes). 
At day 211, no guide stars were available for STIS; however, there were guide stars available for COS. 
To ensure semiconcurrent FUV and NUV observations, we switched our NUV visit to the COS/NUV detector observing with the G230L grating and 2635, 3000, and 3360 central wavelengths to cover the segment gaps and obtain as broad wavelength coverage as possible.

All STIS observations were executed with the 52$\arcsec$x0.2$\arcsec$ slit except for the day 619 visit. 
For the day 619 spectrum, only one guide star was available, introducing the possibility of more drift than usual. 
For this reason, these observations were observed with the 52$\arcsec$x0.5$\arcsec$ aperture.

We acquired our final day 722 spectrum with a blind offset, which resulted in the source being slightly offset in the slit and thus
affected the wavelength solution. 
We fixed this problem following the STScI Cross Correlation notebook\footnote{https://spacetelescope.github.io/hst\_notebooks/notebooks/STIS/cross-correlation/cross-correlation.html}.
Briefly, we recalibrated the data setting the WAVECORR keyword to OMIT, and
then cross-correlated the individual day 722 spectra with the combined day 619 spectra around the strongest features in the G140L and G230L gratings, respectively \Lya\ and \MgII. 
We fit the cross-correlation coefficient as a function of lag (in pixels) to derive the offset in pixels for each observation. 
While the G230L cross-correlation coefficient plot showed a clear peak at $\sim10$ pixels, the G140L was less clear and in all cases consistent with a 0 pixel lag, which we hard-coded as the lag. 
We then used the plate scale to convert the lag in pixels to Angstrom units and updated the SHIFT1 header keyword with these values: $\sim$15.2\,\AA\ for the NUV observations and 0 for the FUV observations.
Finally, we recalibrated the data with \texttt{calstis}.

NUV observations of SN~2024ggi were obtained with the STIS CCD (G230LB) on day 41. 
Before our next intended observation at day 100, HST entered reduced gyro mode operations and SN~2024ggi was no longer schedulable at this phase. 
With this information, and based on the strong detection of Mg~II emission at day 41 and a refinement of the extinction since our proposal submission, we replaced this observation with a visit that included both FUV (G140L) and NUV (G230L) observations at day 232 with the MAMA detector, comparable with the observed phase of SN~2023ixf.
See \autoref{tab:observation} for details regarding the HST observations of SN~2024ggi.
All observations were observed with the 52$\arcsec$x0.2$\arcsec$ slit.

Calibrated NUV 1D spectra were downloaded from MAST and all spectra for a given epoch were combined with the HASP coadd script, as was done for SN~2023ixf.
We remove regions affected by airglow or ISM absorption.
The FUV trace of SN~2024ggi was so faint that the automated extraction of the spectrum failed and we manually extracted it using \texttt{calstis}. 
We optimize the extraction using a 8 pixel extraction box centered on pixel 381. 
To best characterize the background, balancing the signal-to-noise ratio (S/N) in the background region and the background gradient across the detector, we offset the first background region by $-60$ pixels and use a background size of 100 pixels, and offset the second background region by 160 pixels and use a background size of 300 pixels. 
Owing to the strong airglow lines, we turn background smoothing off. 
The background was averaged between the two regions.



\startlongtable
\begin{deluxetable*}{cccccccccc}
\tablecaption{Log of HST UV Observations of SN~2023ixf and SN~2024ggi \label{tab:observation}}
\tablehead{
\colhead{Supernova} &  \colhead{Phase} &\colhead{Average} & \colhead{Grating} & \colhead{Central} & \colhead{Aperture} & \colhead{Instrument} & \colhead{Exposure} & \colhead{Start} & \colhead{Proposal} \\
\colhead{} &  \colhead{(d)} & \colhead{Phase (d)} & \colhead{} & \colhead{Wavelength} & \colhead{} & \colhead{} & \colhead{Time (s)} & \colhead{Time (UT)} &  \colhead{ID}}
\startdata
SN~2023ixf  &183.1 & 199 & G140L & 1425 & 52$\arcsec$X0.2$\arcsec$ & STIS & 1994 & 2023-11-17 20:42:56 &  17497 \\
SN~2023ixf  &183.2 & 199 & G140L & 1425 & 52$\arcsec$X0.2$\arcsec$ & STIS & 2506 & 2023-11-17 22:02:46 &  17497 \\
SN~2023ixf  &183.2 & 199 & G140L & 1425 & 52$\arcsec$X0.2$\arcsec$ & STIS & 2506 & 2023-11-17 23:37:44 &  17497 \\
SN~2023ixf  &183.3 & 199 & G140L & 1425 & 52$\arcsec$X0.2$\arcsec$ & STIS & 2506 & 2023-11-18 01:12:43 &  17497 \\
SN~2023ixf  &214.1 & 199 & G230L & 2635 & PSA                      & COS  & 1700 & 2023-12-18 21:03:37 &  17497 \\
SN~2023ixf  &214.1 & 199 & G230L & 3000 & PSA                      & COS  & 337  & 2023-12-18 21:34:55 &  17497 \\
SN~2023ixf  &214.2 & 199 & G230L & 2950 & PSA                      & COS  & 1700 & 2023-12-18 22:30:02 &  17497 \\
SN~2023ixf  &214.2 & 199 & G230L & 3000 & PSA                      & COS  & 629  & 2023-12-18 23:01:20 &  17497 \\
SN~2023ixf  &214.3 & 199 & G230L & 3360 & PSA                      & COS  & 1700 & 2023-12-19 00:04:58 &  17497 \\
SN~2023ixf  &214.3 & 199 & G230L & 3000 & PSA                      & COS  & 638  & 2023-12-19 00:36:07 &  17497 \\
\hline
SN~2023ixf  &307.8 & 311 & G140L & 1425 & 52$\arcsec$X0.2$\arcsec$ & STIS & 1994 & 2024-03-21 13:24:11 &  17610 \\
SN~2023ixf  &307.9 & 311 & G140L & 1425 & 52$\arcsec$X0.2$\arcsec$ & STIS & 2506 & 2024-03-21 14:48:16 & 17610 \\
SN~2023ixf  &310.7 & 311 & G230L & 2376 & 52$\arcsec$X0.2$\arcsec$ & STIS & 1994 & 2024-03-24 10:55:46 & 17610 \\
SN~2023ixf  &310.8 & 311 & G230L & 2376 & 52$\arcsec$X0.2$\arcsec$ & STIS & 2506 & 2024-03-24 12:22:07 & 17610 \\
SN~2023ixf  &310.8 & 311 & G140L & 1425 & 52$\arcsec$X0.2$\arcsec$ & STIS & 1994 & 2024-03-24 14:05:30 & 17610 \\
SN~2023ixf  &310.9 & 311 & G140L & 1425 & 52$\arcsec$X0.2$\arcsec$ & STIS & 2506 & 2024-03-24 15:31:49 & 17610 \\
SN~2023ixf  &311.6 & 311 & G140L & 1425 & 52$\arcsec$X0.2$\arcsec$ & STIS & 1994 & 2024-03-25 09:07:22 & 17610 \\
SN~2023ixf  &311.7 & 311 & G140L & 1425 & 52$\arcsec$X0.2$\arcsec$ & STIS & 2506 & 2024-03-25 10:29:56 & 17610 \\
SN~2023ixf  &313.6 & 311 & G230L & 2376 & 52$\arcsec$X0.2$\arcsec$ & STIS & 1994 & 2024-03-27 08:29:05 & 17610 \\
SN~2023ixf  &313.7 & 311 & G230L & 2376 & 52$\arcsec$X0.2$\arcsec$ & STIS & 2506 & 2024-03-27 09:55:23 & 17610 \\
\hline
SN~2023ixf  &618.9 & 619 & G230L & 2376 & 52$\arcsec$X0.5$\arcsec$ & STIS & 1934 & 2025-01-26 14:27:10 & 17772 \\
SN~2023ixf  &618.9 & 619 & G230L & 2376 & 52$\arcsec$X0.5$\arcsec$ & STIS & 2670 & 2025-01-26 15:35:06 & 17772 \\
SN~2023ixf  &619.0 & 619 & G140L & 1425 & 52$\arcsec$X0.5$\arcsec$ & STIS & 2628 & 2025-01-26 17:10:18 & 17772 \\
\hline
SN~2023ixf  &722.6 & 722 & G230L & 2376 & 52$\arcsec$X0.2$\arcsec$ & STIS & 2025 & 2025-05-10 09:24:30 & 17772 \\
SN~2023ixf  &722.7 & 722 & G230L & 2376 & 52$\arcsec$X0.2$\arcsec$ & STIS & 2670 & 2025-05-10 10:48:36 & 17772 \\
SN~2023ixf  &722.8 & 722 & G230L & 2376 & 52$\arcsec$X0.2$\arcsec$ & STIS & 2670 & 2025-05-10 12:23:23 & 17772 \\
SN~2023ixf  &722.8 & 722 & G230L & 2376 & 52$\arcsec$X0.2$\arcsec$ & STIS & 2670 & 2025-05-10 13:58:11 & 17772 \\
SN~2023ixf  &722.9 & 722 & G230L & 2376 & 52$\arcsec$X0.2$\arcsec$ & STIS & 2670 & 2025-05-10 15:32:58 & 17772 \\
SN~2023ixf  &723.6 & 722 & G140L & 1425 & 52$\arcsec$X0.2$\arcsec$ & STIS & 2025 & 2025-05-11 09:06:05 & 17772 \\
SN~2023ixf  &723.7 & 722 & G140L & 1425 & 52$\arcsec$X0.2$\arcsec$ & STIS & 2670 & 2025-05-11 10:30:07 & 17772 \\
%
%
\hline
\hline
SN~2024ggi &41.5 & 41  & G230L & 2376 & 52$\arcsec$X0.2$\arcsec$ & STIS & 2243 & 2024-05-22 06:22:55 & 17614 \\
SN~2024ggi &41.5 & 41  & G230L & 2376 & 52$\arcsec$X0.2$\arcsec$ & STIS & 2739 & 2024-05-22 07:49:27 & 17614 \\
\hline
SN~2024ggi &232.1 & 232 & G230L & 2376 & 52$\arcsec$X0.2$\arcsec$ & STIS & 2170 & 2024-11-28 21:24:55 & 17614 \\
SN~2024ggi &232.2 & 232 & G230L & 2376 & 52$\arcsec$X0.2$\arcsec$ & STIS & 2625 & 2024-11-28 22:51:55 & 17614 \\
SN~2024ggi &232.2 & 232 & G140L & 1425 & 52$\arcsec$X0.2$\arcsec$ & STIS & 2625 & 2024-11-29 00:26:29 & 17614 \\
SN~2024ggi &232.3 & 232 & G140L & 1425 & 52$\arcsec$X0.2$\arcsec$ & STIS & 2625 & 2024-11-29 02:01:02 & 17614 \\
\enddata
\end{deluxetable*}

\subsection{Optical Observations}
\begin{figure*}
    \centering
    \includegraphics[width=\textwidth]{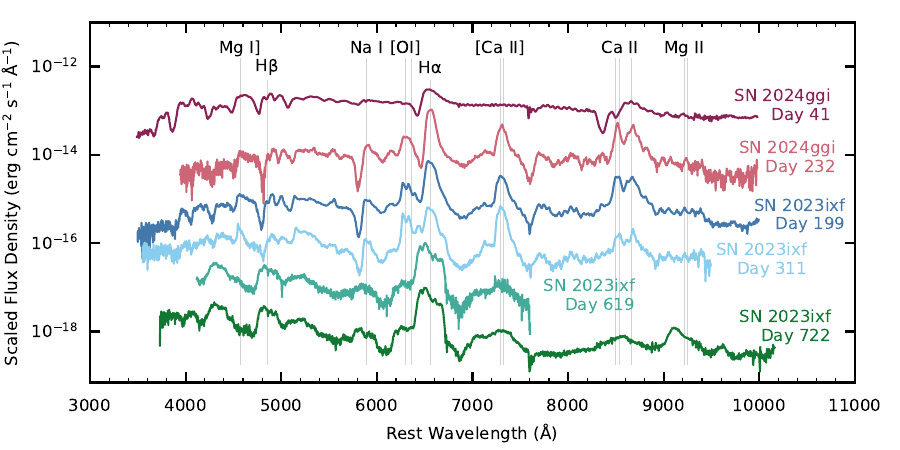}
    \caption{The optical evolution of SN~2024ggi (purple and pink) and SN~2023ixf (blue, light blue, teal, and green). \label{fig:opt}}
    
\end{figure*}

We obtained optical observations at similar epochs to our UV observations; they are listed in \autoref{tab:optical} and shown in \autoref{fig:opt}.
Spectra of SN~2023ixf and SN~2024ggi were obtained with the Las Cumbres Observatory as part of the Global Supernova Project (GSP; PI A. Howell) on  2023-12-03  14:41:23 and 2024-03-19 09:11:22. 
The spectra were taken with the 2\,m Faulkes Telescope North (FTN) in Haleakala, HI, using the FLOYDS spectrograph \citep{2013Brown}.
SN~2024ggi was also observed by the Las Cumbres Observatory as part of the GSP using the FLOYDS spectrograph on FTN (2024-05-25 06:47:34) and the twin instrument on the 2\,m Faulkes Telescope South (FTS) at the Siding Spring Observatory (Australia) on 2024-12-11  16:15:49.
These data were reduced in a standard way with a custom IRAF pipeline \citep{2014Valenti}. 

An additional optical spectrum of SN~2023ixf was obtained with the Binospec instrument \citep{2019Fabricant} at the MMT Observatory as part of the AZTEC collaboration (PI J. Pearson) on 2025-02-05. 
The 2D data reduction (including wavelength and flux calibration) was performed with the Binospec IDL pipeline \citep{2019Kansky}, and the final 1D spectrum was extracted with IRAF \citep{1986Tody, 1993Tody, 2024Fitzpatrick}.

A final optical spectrum of SN~2023ixf was obtained at the Gemini North observatory with the Gemini Multi-Object Spectrograph  \citep[GMOS; ][]{2004Hook, 2016Gimeno} as part of the Distance Less Than 40 Mpc \citep[DLT40; ][]{2018Tartaglia} collaboration (PI. K. A. Bostroem). 
Observations were taken with the B480 grating on 2025-05-17 09:05:55 using the 5200\,\AA\ and 5300\,\AA\ central wavelength settings, and with the R400 grating on 2025-05-18 08:47:33 using the 7650\,\AA\ and 7750\,\AA\ central wavelength positions.
Observations were downloaded from the Gemini Archive and reduced with the Data Reduction for Astronomy from Gemini Observatory North and South \citep[DRAGONS; ][]{2019Labrie} package following the long-slit reduction recipes.

All spectra were scaled to photometric observations taken near the time of the observation by integrating the spectroscopic flux in the observed imaging filters using the \texttt{lightcurve\_fitting} package \citep{LCFitting}.
The first three epochs of SN~2023ixf and all of the SN~2024ggi epochs were scaled to \textit{BVgri} images taken by the Las Cumbres Observatory 0.4\,m and 1\,m telescopes (Hsu et al., in prep.; Wang et al., in prep.). 
Observations were reduced with the Python-based \texttt{BANZAI} pipeline \citep{2018McCully} and photometry was extracted with \texttt{lcogtsnpipe} \citep{2014Valenti}, a PyRAF and Python-based photometry pipeline. 
\textit{B} and \textit{V} zeropoints were calculated using Landolt standard stars, and $g$, $r$, and $i$ zeropoints were calibrated to the Sloan Digital Sky Survey Catalog Data Release 17 \citep{2022Abdurrouf}. 

The final Gemini spectrum was scaled to  $r$- and $i$-band images obtained on 2025-04-21 11:06:45 with GMOS at Gemini-N ($r=19.61$\,mag and $i=20.40$\,mag). 
Given the complex and bright background, the data were reduced using DRAGONS following the imaging recipes, which reduce each CCD detector individually, and photometry was performed using IRAF.
Zeropoints were calculated using the Legacy Survey Catalog \citep{2019Dey}.

\begin{deluxetable*}{cccccc}
\tablecaption{Spectroscopic log for optical observations of SN~2023ixf and SN~2024ggi \label{tab:optical}}
\tablehead{\colhead{Supernova} &  \colhead{Date (UTC)} & \colhead{MJD (day)} & \colhead{Phase (day)} & \colhead{Observatory} & \colhead{Instrument}}
\startdata
    SN~2023ixf & 2023-12-03 14:41:23    & 60281.61 & 199 & Las Cumbres & FLOYDS-N\\
    SN~2023ixf & 2024-03-19 09:11:22    & 60388.38 & 305 & Las Cumbres & FLOYDS-N \\
    SN~2023ixf & 2025-02-05 11:02:51    & 60711.46 & 629 & MMT        & Binospec \\
    SN~2023ixf & 2025-05-17 09:26:42    & 60812.39 & 730 & Gemini-N   & GMOS \\
    \hline
    SN~2024ggi & 2024-05-25 06:47:34    & 60455.28 & 44 & Las Cumbres & FLOYDS-N \\
    SN~2024ggi & 2024-12-11 16:15:49    & 60655.68 & 245 & Las Cumbres & FLOYDS-S\\
\enddata

\end{deluxetable*}

\section{Analysis}\label{sec:analysis}
\subsection{Line Identification and Spectral Evolution}
\begin{figure*}
\centering
    \includegraphics[width=\textwidth]{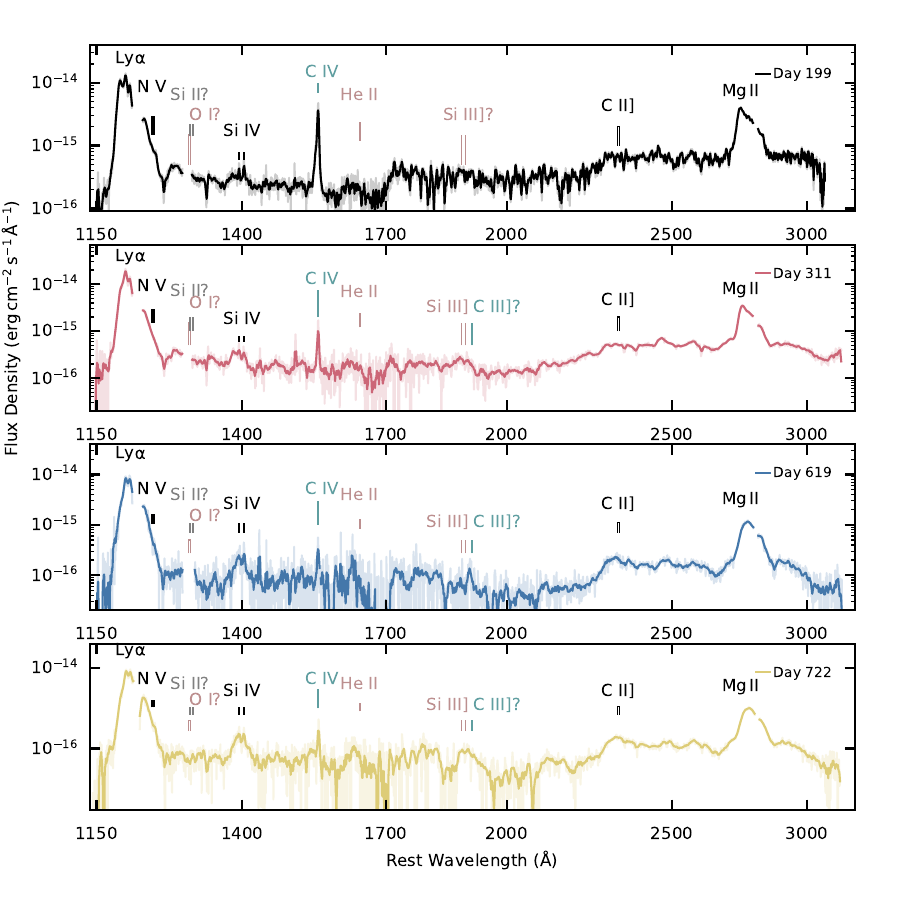}
    \caption{The UV spectroscopic evolution of SN~2023ixf from day 199 to day 722. The most prominent lines are broad, blueshifted \Lya\ and \MgII\ and narrow, unresolved \CIV. 
    Spectra are smoothed to more easily identify features. 
    The unsmoothed spectra are plotted as a semitransparent line in the same color.  
    Regions dominated by airglow emission in \Lya\ and \OIlambda\ and ISM absorption in \MgII\ have been removed.}
    \label{fig:23ixfEvolve}
\end{figure*}
\begin{figure}
    \centering
    \includegraphics[width=\linewidth]{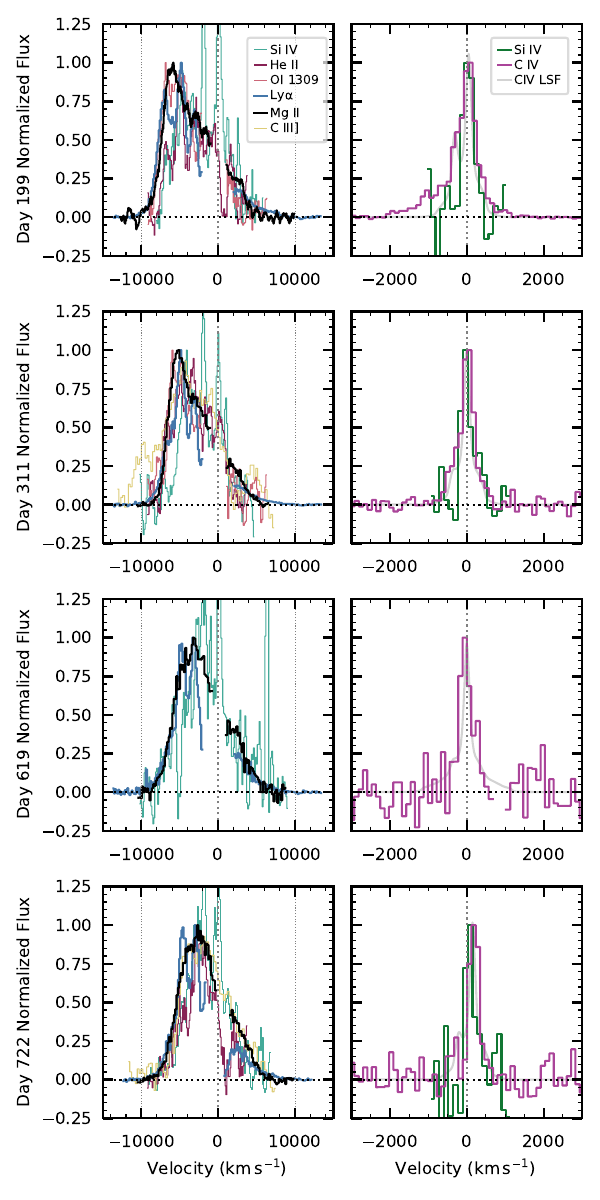}
    \caption{UV line profiles of SN~2023ixf at day 199 (row 1), day 311 (row 2), day 619 (row 3), and day 722 (row 4). \textit{Left}: the broad lines are asymmetric with the blue wings extending to 7500--9000 \kms. 
    The velocity decreases with time and the profiles become less asymmetric (although they are still blueshifted). \textit{Right}: the unresolved, narrow lines are centered on 0 with more symmetric profiles. Both the broad wings characteristic of the HST/STIS line-spread function and the doublet nature of the line are visible in the shape of the line profile. \SiIV\ had unresolved, narrow lines that are superimposed on a broad \SiIV\ which is consistent with the other broad UV features in SN~2023ixf. The line-spread function of the \CIV\ doublet is shown in light gray. }
    \label{fig:uv-profile}
\end{figure}

\autoref{fig:23ixfEvolve} shows the FUV+NUV spectra of SN~2023ixf at days 199, 311, 619, and 722. 
We clearly identify broad and blueshifted \Lya\ and \MgIIlambda. 
\CIVlambda\ is also prominent, although narrow and with the peak centered at 0 velocity in the host reference frame. 
We compare \CIV\ with the STIS line-spread functions for our setups at 1500\,\AA, scaling the flux of each component of the doublet to match our observations, and find that this feature is unresolved (see right-hand panel of \autoref{fig:uv-profile}).
Both \MgII\ and \ion{C}{4} have shallow extended wings on their blue side; it is not possible to tell if red-side wings are present but obscured or nonexistent. 
For \MgII\ these become more prominent with time; however, it is challenging to see if this is also true for \CIV\ as the S/N decreases with time.
The structure in the \Lya\ profile is due to host ISM absorption lines from \ion{Si}{2} $\lambda\lambda\lambda 1190.4$, 1193.3, 1206.5 and \ion{N}{1} $\lambda\lambda 1199.9$, 1200.2. 
These lines were observed in the spectra of SN~2010jl \citep{2014Fransson}.

We also identify broad emission lines from \OIlambda, \SiIVlambda, \HeIIlambda, \SiIIIlambda, \CIIIlambda, and \CIIlambda. 
We note that while we compare the profiles and blueshifts of these lines to the more prominent \MgII\ and \Lya\ to validate their identification, the UV is a complex region dominated by \ion{Fe}{2} absorption and emission which can imitate emission lines from other species.  
There are also narrow lines in the \SiIV\ doublet at rest wavelengths which are also unresolved. 
We do not see any evidence of broad or narrow ionized nitrogen emission lines that have been observed in other interacting supernovae (see \autoref{sec:discussion}).
It may be that the \NVlambda\ emission is lost in the strong \Lya\ profile; however, that does not account for the lack of other nitrogen lines.

The lines identified show temporal evolution. 
\CII, \CIII, \SiIII, and \SiIV\ become more pronounced with time. 
\CIV\ decreases in strength with time relative to the continuum.
The \Lya\ and \MgII\ emission increase in strength with time, relative to the continuum.
Additionally, for the strongest lines (\Lya\ and \MgII), the S/N is high enough to observe the peak flux shift redward and the whole profile become more symmetric (although still blueshifted) with time. We quantify this evolution by measuring the equivalent width and total flux of the lines over time in \autoref{tab:EW}.

\begin{figure*}
    \centering
    \includegraphics[width=\textwidth]{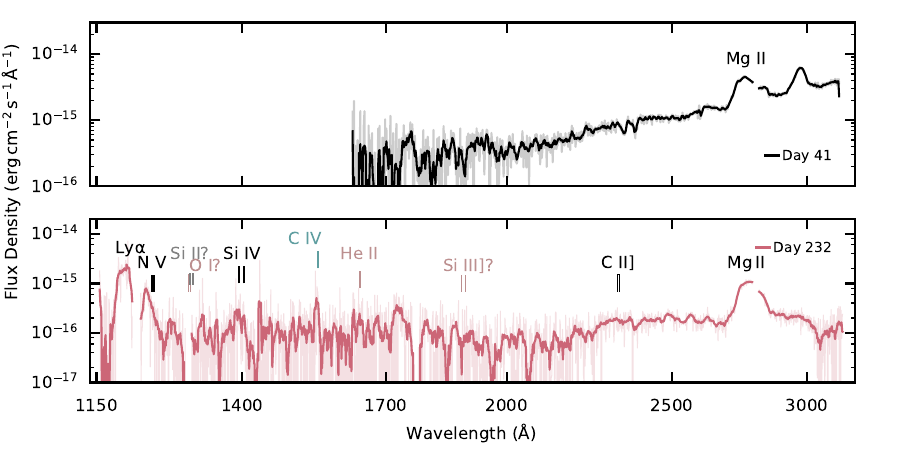}
    \caption{The UV spectroscopic evolution of SN~2024ggi at days 41 and 232. The most prominent lines are broad, blueshifted \Lya\ and narrow, unresolved \CIV\ at day 232 and broad, blueshifted \MgII\ in both epochs. Regions dominated by airglow emission in \Lya\ and \OIlambda\ and ISM absorption in \MgII\ have been removed. \label{fig:24ggiTimeEvolve}}
    
\end{figure*}

The UV spectra of SN~2024ggi are shown in \autoref{fig:24ggiTimeEvolve} at days 41 and  232.
The NUV spectrum of SN~2024ggi at day 41 exhibits a clear, asymmetric, and blueshifted \MgII\ feature. 
Additionally, there is another prominent emission feature at $\sim$2973\,\AA\ which is not present in the day 66 SN~2023ixf spectrum. 
The shape of this feature is much narrower and more symmetric than \MgII\, indicating that its peak may be at its rest wavelength. 
The closest line to this location would be \ion{Fe}{1} $\lambda\lambda 2973$, 2981. 
Identification of features in the UV is challenging as metal absorption lines overlap. 
It is also possible that this is not a line at all and that it represents a window of low metal absorption.
\citet{2023Bostroem1} found that models of the spectrum of SN~2022acko at day 20 showed a region of low absorption between nearby \ion{Fe}{2}, \ion{Cr}{2}, and \ion{Ti}{2} absorption complexes which looked like an emission feature at 2970 \AA. 

The FUV+NUV spectrum of SN~2024ggi at day 232 is very similar to that of SN~2023ixf at day 199, showing broad, asymmetric, and blueshifted \Lya\ and \MgII, and narrow, unresolved \ion{C}{4}. 
The rise to a plateau between \ion{C}{2}] and \MgII\ also matches that seen in SN~2023ixf.
The extinction-corrected UV flux of SN~2024ggi is a factor of 4 smaller than that of SN~2023ixf at this phase, and the low S/N makes it more challenging to identify features. 
We tentatively identify blueshifted \OIlambda. 
The only other clearly identifiable feature is broad \ion{Si}{4}, which is stronger than that of SN~2023ixf at a similar phase.

At day 41, the flux and equivalent width of \MgII\ is lower in SN~2024ggi than in SN~2023ixf. 
By day $\sim$200 the \MgII\ line in SN~2024ggi has a smaller flux but larger equivalent width than in SN~2023ixf. 
The flux and equivalent width of \Lya\ are also smaller for SN~2024ggi than for SN~2023ixf at day $\sim$200. 
Comparing the profiles of SN~2023ixf and SN~2024ggi on both days $\sim$50 and $\sim$200, SN~2024ggi is more symmetric, indicating a more symmetric CSM or less occultation of the far side of the ejecta and CSM. 
In SN~2024ggi, high-ionization optical emission disappeared much faster than in SN~2023ixf \citep{2024Shrestha}. 
Assuming both SN~2024ggi and SN~2023ixf had similar ejecta velocities, this implies a smaller radial extent of the CSM. 
If additionally, the CSM around both SN~2023ixf and SN~2024ggi had the same density, then we would expect a smaller mass to be swept up in the CDS of SN~2024ggi. 
Observationally, this could manifest as a more symmetric emission line (as the CDS would have a lower optical depth) and a higher maximum velocity (as the outer ejecta would not have been slowed down as much by the formation of the CDS). 
Differences in the progenitor, CSM, and explosion properties could also produce different CDS properties.

\begin{table*}
        \caption{The equivalent width and flux of the strongest emission lines in SN~2023ixf and SN~2024ggi \label{tab:EW}}
    \centering
    \begin{tabular}{|c|c|c|c|c|c|c|c|}
    \hline
     && \multicolumn{3}{|c|}{Equivalent Width (\AA)} &  \multicolumn{3}{|c|}{Flux ($\mathrm{erg~cm^{-2}~s^{-1}}$)} \\
    \hline
     \multirow{2}{*}{Supernova} & \multirow{2}{*}{Phase (d)}         & \multirow{2}{*}{\Lya} & \CIV\ & \MgII\  & \multirow{2}{*}{\Lya} & \CIV\ & \MgII\ \\
      &                    &  & $\lambda\lambda1548.9, 1550.8$ & $\lambda\lambda2795.5, 2802.7$ &             &  $\lambda\lambda1548.9, 1550.8$ & $\lambda\lambda2795.5, 2802.7$ \\
    \hline
    SN~2023ixf &  66 &   -    &   -    & $-$239.9  &   -    &   -    & $3.67\times10^{-13}$   \\
    SN~2023ixf & 199 & $-$1625.6 & $-$95.0 & $-$325.9  & $2.15\times10^{-13}$ & $1.39\times10^{-14}$ & $1.72\times10^{-13}$ \\
    SN~2023ixf & 311 & $-$1984.1 & $-$27.7 & $-$361.1  & $2.52\times10^{-13}$ & $3.11\times10^{-15}$ & $1.43\times10^{-13}$ \\
    SN~2023ixf & 619 & $-$2867.5  & $-$13.7 & $-$459.5 & $1.32\times10^{-13}$  & $7.25\times10^{-16}$ & $5.65\times10^{-13}$\\
    SN~2023ixf & 722 & $-$4950.9  & $-$23.8 & $-$526.3 & $1.26\times10^{-13}$  & $4.20\times10^{-16}$ & $4.99\times10^{-14}$\\
    \hline
    SN~2024ggi & 41 &- & -&  $-$116.8       &- & -& $9.37\times10^{-14}$\\
    SN~2024ggi & 232 & $-$937.9 & - & $-$404.7 & $1.18\times10^{-14}$ & - & $3.24\times10^{-14}$\\
    \hline
    
    \end{tabular}

\end{table*}


\subsection{Velocity Comparison and Evolution}
The velocity of a line can be mapped to different regions in the ejecta and surrounding material. 
In supernovae, broad lines with maximum velocities of order 1000--10,000 \kms\ originate from the supernova ejecta and/or CSM that has been shocked by the ejecta.
In a spherically symmetric, expanding, optically thin medium, a symmetric emission line will form around 0 velocity with the maximum red and blue-side velocities representing the maximum velocity of the emitting material. 
When there is some optical depth to the medium, the photons from the back side of the ejecta can be obscured, attenuating the red side of the line profile. 
In this scenario, the blue side of the profile is coming from photons emitted toward the observer and therefore is not absorbed. 
Assuming homologous expansion, material at a larger radius is expanding at a higher velocity. 
Therefore, when two lines are emitted from two different regions (with different radii), the material emitted at the larger radius will have broader features (and a higher maximum velocity) which is most reliably measured from the blue side of the line profile. 
Furthermore, unshocked CSM is moving at the stellar wind speed (of order 10--100 \kms). 
This material is expected to produce narrow lines, like those found in Type IIn supernovae. 
We measure the velocity and temporal evolution of UV and optical emission lines in SN~2023ixf and SN~2024ggi and use these to build a physical picture of the supernova ejecta and CSM.  

We start by comparing the UV line profiles of SN~2023ixf to definitively identify broad and narrow features in \autoref{fig:uv-profile}. 
To compare the line profiles, we subtract a linear continuum, fit to a small region on either side of the line, and normalize the maximum flux in the continuum-subtracted region.
If the normalization is dominated by noise spikes, we smooth the spectrum to find the normalization.
We then calculate the velocity using the rest wavelength of the line identified. 
In cases where a line is a doublet, we use the first line in the doublet, except for the narrow \CIV, for which we find the peak better aligns with 1550.8\,\AA.
For the strongest lines, we calculate the velocity at which 99\% of the flux is included using the smoothed spectrum and report both the blue- and red-side velocity in \autoref{tab:velocity}. 
This flux depends sensitively on both the choice of continuum and integration limits. 
We vary the continuum choice by $\pm$1000 \kms\ and integrate the smoothed and unsmoothed profile to characterize some of this uncertainty. 
The largest deviation is reported in the fourth column of \autoref{tab:velocity}.
Starting at day 311, there is a low level of flux in the blue wing of \MgIIlambda. 
From this flux, the \MgII\ line profile rises steeply at velocities consistent with other UV lines.
We therefore consider this contamination from an alternate species and do not consider this wing in this analysis.

At day 200, the broad lines (\Lya, \OIlambda, \SiIVlambda, \HeIIlambda, \CIIIlambda, \MgII)  are all blueshifted with a maximum velocity (from the blue edge of the profile) of $\sim$7500--9000 \kms\ while the red edge lies at 3000--4500 \kms. 
 \Lya{}, \OI\, and \MgII\ are very asymmetric, while the other lines appear to be more symmetric and at a slightly lower maximum velocity, although the S/N is lower. 
Over time, the emission becomes less asymmetric (although still blueshifted) and there is a more uniform profile shape across all UV lines. 
Additionally, the blue-side maximum velocity decreases with time to about 6000--7000 \kms\ and the red side increases to 4000--5500 \kms\ at day 722. 

Prominent narrow lines are present in \SiIV\  and \CIVlambda. 
The line profiles are mostly symmetric around 0 \kms\ and are unresolved (the full width at half-maximum intensity [FWHM] of a resolution element at this wavelength is 170 \kms --- but note that the HST/STIS line-spread function is not a Gaussian). 
Given the width of these lines, they cannot originate from the supernova ejecta or the CDS and must therefore be from the CSM.

\begin{figure}
    \centering
    \includegraphics[width=\columnwidth]{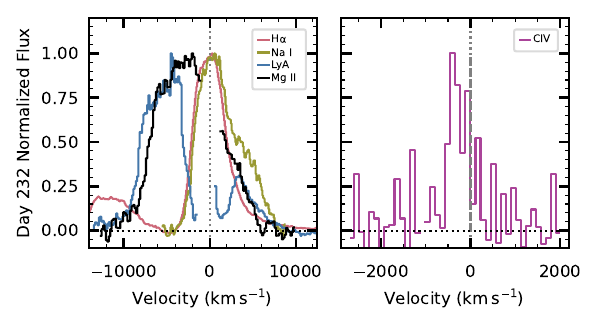}
    \caption{\textit{Left}: A comparison of the broad UV and optical line profiles at day 232 for SN~2024ggi. 
    The blue edge of the UV lines, \Lya\ and \MgIIlambda, is at a significantly higher velocity than the optical lines, \Ha\ and \ion{Na}{1}, indicating that these are originating from distinct parts of the ejecta/CSM. 
    \textit{Right}: A third velocity component is seen in the unresolved, narrow \CIVlambda\ emission line from the CSM which has not been accelerated by the ejecta.  0 velocity and flux are marked with dotted lines.} 
    \label{fig:uv-opt-profile-24ggi}
\end{figure}

The \Lya, \CIV, and \MgII\ lines profiles for SN~2024ggi at day 232 are shown in \autoref{fig:uv-opt-profile-24ggi} and their 99\% velocities are reported in \autoref{tab:velocity}. 
Like SN~2023ixf, the \Lya\ and \MgII\ profiles are nearly coincident with the blue wings extending to $\sim$9000 \kms. 
The line profiles are also blueshifted and asymmetric --- although less extreme than in SN~2023ixf.
Narrow, unresolved \CIV\ is also present at lower S/N with lower velocities than is seen at 199\,d for SN~2023ixf.

\begin{table}
\centering
\caption{The velocity containing 99\% of the flux of the most prominent lines for SN~2023ixf and SN~2024ggi. 
We vary the continuum region and smoothing and report the maximum deviation in Column 4 with a minimum uncertainty of 100 \kms.\label{tab:velocity}}
\begin{tabular}{|c|c|c|c|c|}
\hline
\multirow{2}{*}{Species}  & $\mathrm{V_{max, blue}}$ & $\mathrm{V_{max, red}}$ & Uncertainty\tablenote{This is not a standard deviation} & Spectral \\
 & (\kms) & (\kms) & (\kms) &  Region\\
\hline
\multicolumn{5}{|c|}{SN~2024ggi 232d}\\
\hline
\MgII\ & $-$8850 & 5350 & 150 & UV\\
\ion{Na}{1} & $-$3150 & 7400 & 950 & Optical\\
\hline
\multicolumn{5}{|c|}{SN~2023ixf 199d}\\
\hline
\ion{Si}{4} & $-$7450 & 2750 & 1100 & UV\\
\MgII\ & $-$9000 & 4400 & 250 & UV\\
\ion{Na}{1} & $-$2950 & 6450 & 650 & Optical\\
\hline
\multicolumn{5}{|c|}{SN~2023ixf 311d}\\
\hline
\ion{Si}{4} & $-$5900 & 2700 & 350 & UV \\
\MgII\ & $-$7900 & 4550 & 650 & UV\\
\ion{Na}{1} & $-$2550 & 5700 & 350 & Optical\\
\hline
\multicolumn{5}{|c|}{SN~2023ixf 619d}\\
\hline
\ion{Si}{4} & $-$7750 & 4500 & 150 & UV \\
\MgII\ & $-$8200 & 4800 & 350 & UV\\
\Ha\ & $-$8300 & 7100 & 100 & Optical\\
\hline
\multicolumn{5}{|c|}{SN~2023ixf 722d}\\
\hline
\ion{Si}{4} & $-$6600 & 4250 & 1250 & UV \\
\MgII\ & $-$7200 & 5450 & 650 & UV\\
\Ha\ & $-$8150 & 6950 & 100 & Optical\\
\hline
\end{tabular}

\end{table}

\begin{figure}
    \centering
    \includegraphics[width=\linewidth]{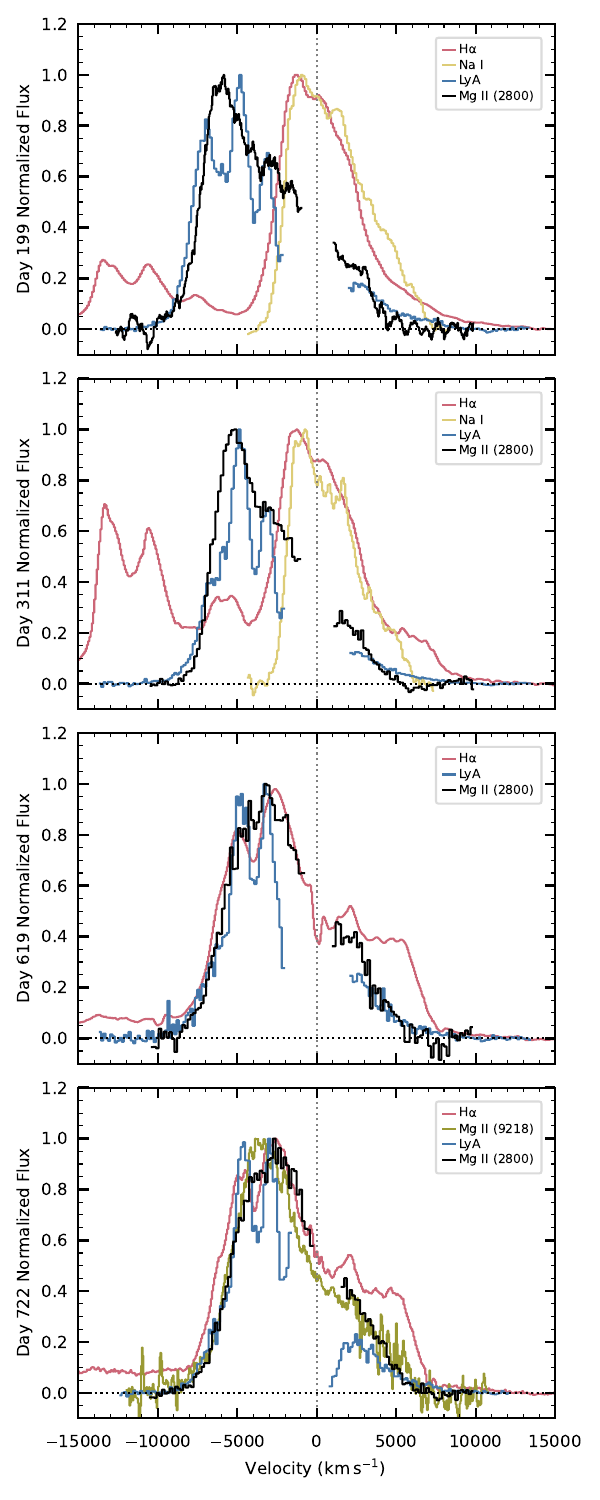}
    \caption{A comparison of the UV and optical lines profiles of SN~2023ixf  at day 199 (row 1), day 311 (row 2), day 619 (row 3), and day 722 (row 4). At early phases the UV and optical lines show distinct profiles with the UV being blueshifted and asymmetric while the optical lines are centered around 0 velocity with narrower widths. After day 611, the optical line profiles evolve to match the UV lines, indicating that CSM interaction is now the dominant power source.}
    \label{fig:uv-opt-profile}
\end{figure}

When the supernova ejecta enter the recombination phase, the ejecta flux shifts into the optical and the UV flux fades.
If CSM is present, as CSM interaction becomes the dominant energy source powering the outer ejecta, the spectrum becomes dominated by broad lines which overpower the subdominant radioactive decay that heats the inner, metal-rich ejecta and produces relatively narrower lines \citep{2022Dessart, 2023Dessart}. 
To investigate this scenario, we compare the UV line profiles to \Ha\ and, when present, \ion{Na}{1} $\lambda \lambda 5890$, 5896 and \ion{Mg}{2} $\lambda\lambda9218$, 9244 for SN~2023ixf (\autoref{fig:uv-opt-profile}) and SN~2024ggi (\autoref{fig:uv-opt-profile-24ggi}). 
At day $\sim200$, the UV and optical lines of both supernovae are clearly separated with the optical lines centered at 0 velocity. 
This is consistent with the optical lines originating from the inner ejecta and the UV lines originating from the CDS.
\Ha\ has an extended red wing reminiscent of that seen on the blue side of \MgII\ in SN~2023ixf.
\Lya\ and \MgII\ are coincident and blueshifted in the UV.
In SN~2023ixf, the blue wings of \Ha{} extend slightly further to around 5000 \kms\ while \ion{Na}{1} extend to 4000 \kms. 
In SN~2024ggi, these lines are coincident on the blue side with a maximum velocity slightly below 5000 \kms.

For SN~2023ixf, we can assess the temporal evolution of these features. 
Similar profiles are seen at day 311. 
The \Lya\ and \MgII\ lines are coincident, broad, and blueshifted with a maximum velocity around 8000 \kms. 
The optical lines are narrower, with a maximum velocity around 2500 \kms\ and centered at 0 \kms. 
At this phase a new emission feature is more prominent blueward of \Ha\ with a peak velocity that matches the peak velocity of the UV lines at $\sim$5000 \kms. 
On the red side of \Ha\ there is a weaker emission component at a similar velocity to the blue emission feature.
The high velocity components have been noted by others in earlier optical spectra as early as 84 days after explosion \citep{2024Ferrari, 2025Folatelli,2025Zheng,2025Kumar}.
At day 619, a dramatic shift has occurred and \Ha{} is now coincident with the UV lines, \Lya\ and \MgII{}. 
All three lines have a maximum velocity of around 8200 \kms\ and are blueshifted with identical blue-side profiles and similar asymmetric red sides. 
The boxy emission on the red side of \Ha\ continues to be present here which does not match the UV line profiles. 
This is likely due to the CDS having a lower optical depth in the optical.
Day 722 is similar to day 619.
The spectra obtained at this phase include \ion{Mg}{2} $\lambda\lambda9218$, 9244, a line only expected to originate from CSM interaction \citep{2023Dessart} and which is also identical to the \Lya\ and \MgIIlambda\ line profiles. 

\section{Comparison to Models} \label{sec:models}
The time-dependent non-local thermodynamic equilibrium (NLTE) radiative transfer code \texttt{CMFGEN} \citep{2012Hillier} has been used to model UV spectra in the presence of strong and weak CSM interaction and in the presence of dust \citep{2017Dessart, 2022Dessart, 2023Dessart, 2025Dessart}. 
These models perform the radiative transfer consistently across the full FUV to far IR wavelength range.
Our time series presents a unique opportunity to test how well these models match observations in the UV and optical --- validating the modeling assumptions and indicating new directions.
In particular, even for supernovae with low-density CSM, the UV is expected to be dominated by ejecta-CSM interaction during and after the hydrogen recombination phase. 
Additionally, given the sensitivity of UV wavelengths to dust, we can use these spectra to search for the presence of dust and to understand its location in the ejecta and CSM \citep{2025Dessart}. 
We compare the UV+optical spectra of SN~2023ixf and SN~2024ggi to these models, focusing on the UV region.

\subsection{SN~2024ggi}
\begin{figure*}
\centering
    \includegraphics[width=\textwidth]{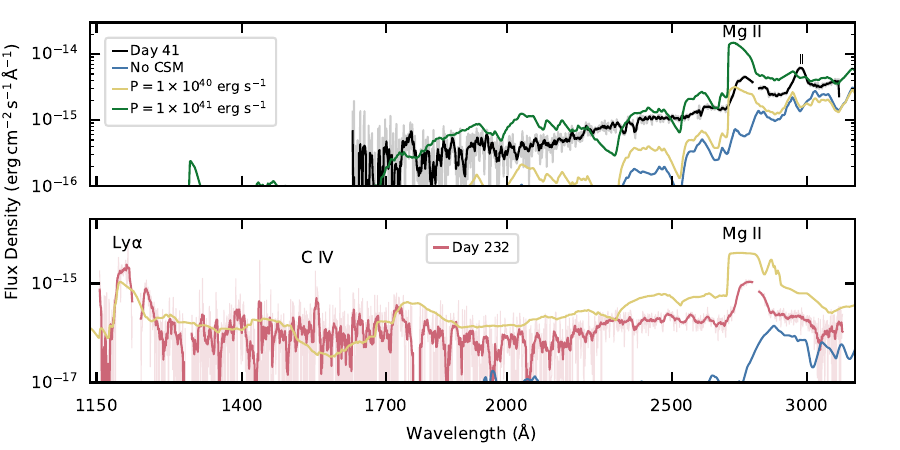}
    \caption{A comparison of the observed spectrum of SN~2024ggi at days 41 (top) and 232 (bottom) to the models of \citet{2022Dessart}. 
    Both phases are inconsistent with the model with no CSM, with the best-matched model at day 41 having a shock power of $P_{\mathrm{shock,abs}}=1\times10^{41}$ \pwrunit\ and the best-matched model at day 232 having a shock power of $P_{\mathrm{shock,abs}}=1\times10^{40}$ \pwrunit.
    }
    \label{fig:model232d}
\end{figure*}

We start by comparing the days 41 and 232 spectra of SN~2024ggi to models at corresponding phases from \citet{2022Dessart} in \autoref{fig:model232d}.
Briefly, these models were created with \texttt{CMFGEN} using a 15 \msun\ RSG progenitor evolved in Modules for Experiments in Stellar Astrophysics \citep{2011Paxton,2013Paxton} and exploded by \texttt{V1D} \citep{1993Livne,2010Dessart1} as described by \citet{2021Dessart}.
Chemical mixing is simulated using boxcar smoothing and a $\mathrm{^{56}Ni}$ mass of $M_{\rm Ni}=0.063$ \msun\ is used for the radioactive decay, within the range of values derived for SN~2023ixf \citep[0.05--0.07 \msun; Jacobson-Galan et al. in prep.;][]{2023Zimmerman, 2024Singh, 2025Li, 2025Vinko}. 
In \citet{2022Dessart}, these models are modified to add a 0.1 \msun\ region of higher density in the outer ejecta at 11700 \kms\ which approximates the CDS and power is injected at this location, mimicking the reprocessing of X-ray by the CDS. 
We emphasize that if the X-rays are not fully thermalized by the CDS, then this will represent a fraction of the total X-ray flux (see \autoref{sec:discussion} for more details).
We consider absorbed shock power values of $P_{\mathrm{shock,abs}}=1\times10^{40}$ \pwrunit, $P_{\mathrm{shock,abs}}=1\times10^{41}$ \pwrunit, $P_{\mathrm{shock,abs}}=5\times10^{41}$ \pwrunit, and $P_{\mathrm{shock,abs}}=1\times10^{42}$ \pwrunit.
In these models, the CDS was placed at 11700 \kms\ to model normal Type IIP/L (i.e. those without early, transient, narrow features). 
In this scenario it is assumed that there is enough CSM to create a CDS but that this does not significantly slow down the outer ejecta and CDS.
The narrow lines in the early optical spectra of SN~2024ggi imply a region of higher CSM density which is predicted to reduce the velocity of the CDS. 
Thus, we expect the velocity of features that originate in the CDS to be overestimated by the model. 

At day 41, the NUV flux is best matched by the model with $P_{\mathrm{shock,abs}}=1\times10^{41}$ \pwrunit.
The flux of \MgIIlambda\, the most prominent line, is overestimated by the $P_{\mathrm{shock,abs}}=1\times10^{41}$ \pwrunit\ model and the flux is better matched by the $P_{\mathrm{shock,abs}}=1\times10^{40}$ \pwrunit\ model. 
The shape of the \MgII\ line is also considerably different; the models are boxier with steeply rising sides and less attenuation of the red side of the profile. 
The shape of the feature is determined by the optical depth and geometry of the emitting region. 
In these models, this line originates in the CDS and the mismatch in shape could be due to the width of the CDS in the models, the simplistic Gaussian density profile in the 1D models or the amount of ``clumping'' added to the CDS which permits more photons to escape. 
The feature around 2970\AA\ is not present in the model.
Although a feature exists in the models at a slightly higher velocity, it exists in the absence of CSM interaction and thus is unlikely to be blueshifted.
The optical is well represented by all of the models with the absorption depth of P Cygni profiles best matched by the $P_{\mathrm{shock,abs}}=1\times10^{41}$ \pwrunit\ model.
These spectra demonstrate the ability of the UV wavelengths to distinguish models with varying power levels which are very similar in the optical.

At day 232 the FUV+NUV observation is best matched by the $P=1\times10^{40}$ \pwrunit\ model.
While this model overestimates the NUV flux, it is significantly better than the model without CSM.
The model underpredicts the \Lya\ flux and overpredicts the \MgII\ flux. 
Both of these are resonance lines and thus strongly affected by optical depth. 
It is possible that this could be rectified by a small change in the density of the CDS in the model. 
The narrow \CIV\ line is not present in the model as it originates from unshocked CSM that is not part of the model. 
Weaker individual lines are not discernible in the model --- possibly a result of the high velocity overlapping the features.
Including the optical does not provide additional information; all models match the continuum well and overestimate many absorption features.

\subsection{SN~2023ixf}
\begin{figure*}
\centering
    \includegraphics[width=1\textwidth]{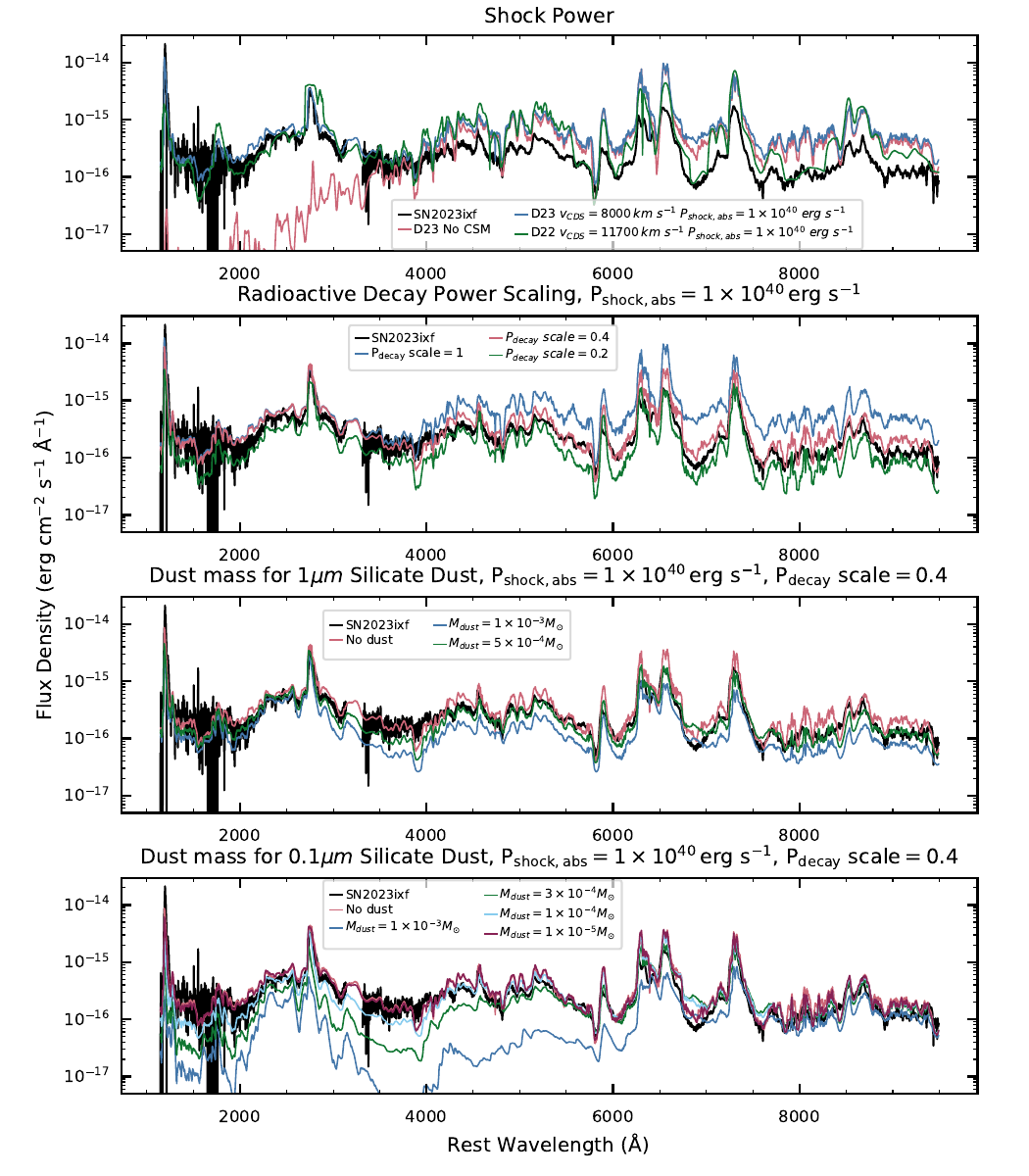}
    \caption{ The effects of shock power (from CSM interaction; first panel), scaling the radioactive decay power (second panel), adding different masses of 1\micron\ silicate dust (third panel), and adding different masses of 0.1\micron\ of silicate dust (fourth panel). While shock power primarily affects UV wavelengths,  radioactive decay primarily affects optical wavelengths. The 1\micron\ dust has very little affect on the UV while the 0.1\micron\ dust can cause significant UV absorption for the highest masses examined. Both dust grain sizes suppress the optical flux, especially the emission lines. \label{fig:all_model}}
\end{figure*}

We created custom models of SN~2023ixf with \texttt{CMFGEN} using a 15.2 \msun\ RSG progenitor from \citet{2016Sukhbold} and chemical mixing with the shuffling method as \citet{2023Dessart}.
Following \citet{2023Dessart}, the CDS is placed at $v=8000$ \kms\ with a mass that is twice the ejecta mass above $v=8000$ \kms.
We explore three values of absorbed shock power: $P_{\mathrm{shock,abs}}=7\times10^{40}$ \pwrunit, $P_{\mathrm{shock,abs}}=1\times10^{40}$ \pwrunit, and $P_{\mathrm{shock,abs}}=5\times10^{39}$ \pwrunit. 
Comparing the model at 350\,d to the 311\,d spectrum of SN~2023ixf, we find that the $P_{\mathrm{shock,abs}}=7\times10^{40}$ \pwrunit\ model clearly overestimates the UV flux.
The two lower shock power models match the NUV observations equally well, with  $P_{\mathrm{shock,abs}}=1\times10^{40}$ \pwrunit\ matching the FUV better than  $P_{\mathrm{shock,abs}}=5\times10^{39}$ \pwrunit. 
We compare the spectrum at 311d to the 350d model with no shock power from CSM interaction, $P_{\mathrm{shock,abs}}=1\times10^{40}$ \pwrunit\ from \citet{2022Dessart}, and the $P_{\mathrm{shock,abs}}=1\times10^{40}$ \pwrunit\ custom model in the first panel of \autoref{fig:all_model}.
The optical wavelengths fade by 0.6 mag between day 305 when the optical spectrum was taken and day 350. We scale the optical flux by this amount.
There is no UV photometry during this phase and so we do not attempt to scale the UV spectrum although as this is dominated by CSM interaction, we do no expect the effect to be as significant. 

It is clear that the while all three models are similar in the optical, the model without shock power does not produce enough UV flux. This is rectified by the models with $P_{\mathrm{shock,abs}}=1\times10^{40}$ \pwrunit\ both of which reproduce the UV flux well but over-estimate the optical flux.
As anticipated, the lower CDS velocity of \citet{2023Dessart} more closely matches the observed UV line profiles (this is particularly noticeable in \MgII). 

\autoref{fig:all_model} demonstrates the power of the large wavelength range covered by the UV+optical spectra to tightly constrain the SED, which in turn can constrain the physics. For instance, early UV+optical spectra were used to constrain the ejecta clumping in SN~2022acko \citep{2023Bostroem1}. 
In the presence of ejecta-CSM interaction and at late times, the UV+optical spectra can be used to disentangle the two primary power sources.
The UV constrains the ejecta-CSM interaction (first panel of \autoref{fig:all_model}) and the optical constrains the radioactive decay power of $\rm{^{56}Co}$ absorbed by the ejecta (second panel of \autoref{fig:all_model}).  
In addition to the direct model, we scale the radioactive decay power by 20\% and 40\%. 
All three scalings are shown in the second panel of \autoref{fig:all_model}.
The radioactive decay power scaling that best matches the UV+optical observed spectrum is 0.4.
While our models have a similar $\rm{^{56}Co}$ mass, it is possible that the model underestimates the escape fraction of gamma rays and this scaling factor compensates for that.
The short plateau length of SN~2023ixf \citep{2023Zimmerman, 2024Yang, 2024Hsu, 2024Moriya,2024Singh,2025Zheng,2025Li, 2025Michel} implies a small ejecta mass, which would increase the gamma-ray escape fraction.

Finally, we explore the best-matching dust parameters with the module \texttt{CMF\_FLUX} \citep{2025Dessart}. 
We show models with different dust masses using 1 \micron\ silicate dust in the third panel of \autoref{fig:all_model} and models that use 0.1 \micron\ in the bottom panel. 
We consider 1\micron\ dust masses of $M = 5\times10^{-4}$ \msun\ and $M = 1\times10^{-3}$ \msun\ and 0.1\micron\ dust masses of $M = 1\times10^{-5}$ \msun, $M = 1\times10^{-4}$ \msun, $M = 3\times10^{-4}$ \msun, and $M = 1\times10^{-3}$ \msun. 
We place the dust in the CDS at $v=8000$ \kms\ to better model the conditions in SN~2023ixf \citep{2025Dessart} and set the dust temperature to 650K (see Jacobson-Galan et al. in prep).
Overall, 1 \micron\ dust minimally affects the UV, mildly suppressing the optical continuum and significantly decreasing the optical emission line flux. 
A zoom in on the more prominent UV and optical emission lines is shown in \autoref{fig:ModelLineFlux}.
The 0.1\micron\ dust affects both the UV and optical for all but the lowest mass model. 
While we can rule out the higher mass 0.1\micron\ models, we cannot differentiate between the different dust grain sizes. 
We refer the reader to a more detailed discussion of the effects of dust at longer wavelengths by Jacobson-Galan et al. (in prep.) and \citet{2025Medler}.

Although counterintuitive to the effects of dust, the dust more strongly affects the optical than the UV. 
\citet{2025Dessart} suggest that this could be due to the UV radiation originating at the outer edge of the CDS while the dust is concentrated on the densest regions of the CDS. At the same time, the optical originates from the inner ejecta or denser regions of the CDS, closer to or interior to the dust. 


To examine how these parameters evolve over time, we compare our day 619 spectra to models at day 600. 
At this phase the power from ejecta-CSM interaction has decreased and the observations are best matched by models with a shock power of $P_{shock, abs}=5\times10^{39}$\,\pwrunit. 
We find the UV+optical observations are best matched by a dust mass of $M=1\times10^{-3}$ \msun\ with dust temperature of $T=600$\,K. 
We note that this still slightly overestimates the NUV flux, although less than most of the other models, possibly indicating that the shock power is too large.

UV spectra of supernovae are a combination of broad absorption and emission lines that blend together, making it challenging to definitely identify individual emission and absorption features. 
To identify the effect of an individual species (element+ionization), we run the model spectra without that species. 
We then compare this to the full spectrum to identify the features that are only present when the species is included, which we associate with the element and ionization level. 
For this exercise, we use the best models without dust, as only models in which dust has a minimal effect on UV were consistent with our observations (see \autoref{fig:all_model}). 

\begin{figure*}
\centering
    \includegraphics[width=\textwidth]{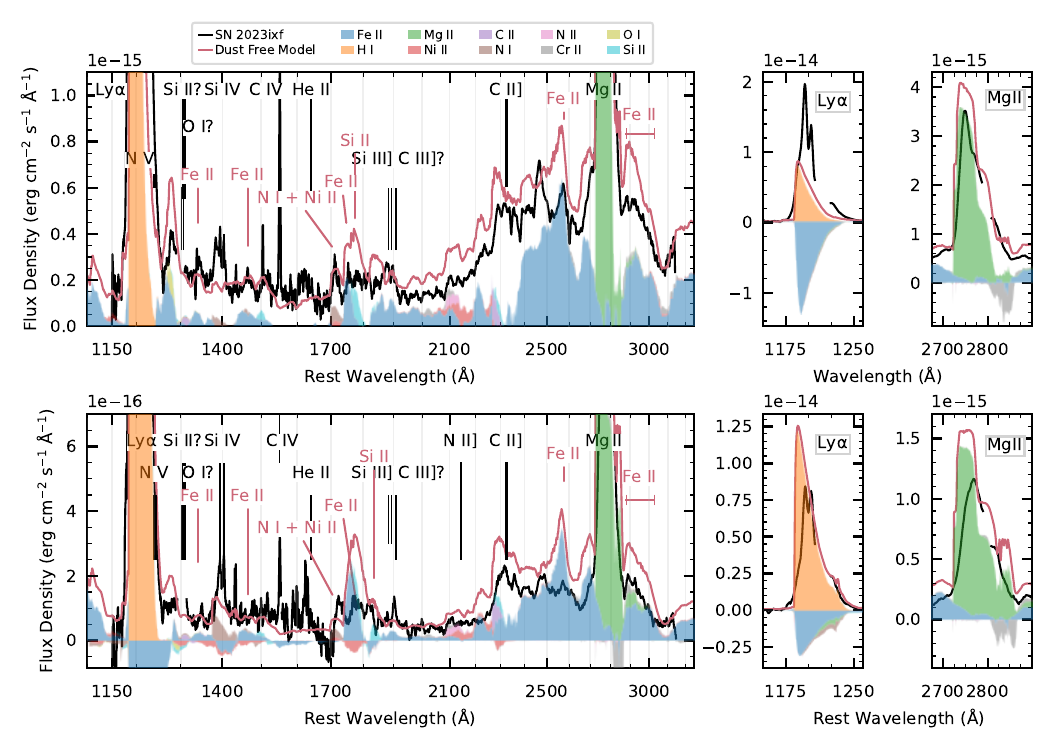}
    \caption{A comparison of the best-matched model without dust at 350\,d (top) and 600\,d (bottom) with the observed UV spectrum at 311\,d and 619\,d. 
    The contributions from individual species are shown as shaded regions and features identified from the models are labeled in pink.
    \textit{Left}: The full model spectra without dust (pink) are compared to the observed spectrum (black) and species identified from the model spectra are marked in pink.
    \textit{Center}: A zoom-in on the \Lya\ region showing the observed flux (black), the full model flux (pink), and contributions from individual elements, dominated by \ion{H}{1} (orange) and \ion{Fe}{2} (blue).
    \textit{Right}: A zoom-in on the \MgII\ region showing the observed flux (black), the full model flux (pink), and contributions from individual elements, dominated by \MgII\ (green), \ion{Fe}{2} (blue) on the wings, and \ion{Cr}{2} (gray) in absorption.
    Only species that produce a noticeable contribution to the flux are shown.}
    \label{fig:modelElement}
\end{figure*}

While we initially envisioned this as a way to verify the elements we identified in \autoref{fig:23ixfEvolve}, the ions of many of the high-ionization states we identify are not included in the model.
Notably these include \SiIV,  \SiIII, \CIV, and \CIII. 
Nevertheless, the overall spectrum is well reproduced, indicating that the broad physics in the model is correct (see \autoref{fig:modelElement}).
The fact that the overall spectrum is well modeled without the high-ionization species implies that their contribution to the overall spectrum is low. 
This could be because the observed features are formed in low-density material which represents a small fraction of the emitting region.
The unresolved, narrow lines likely arise from unshocked CSM which is also not part of the model and is not expected to contribute to the overall spectrum. 

We use the models to confirm lower ionization species. 
\autoref{fig:modelElement}) shows the contribution from individual elements as well as the total model compared to the observed spectrum. 
The models verify the identification of the dominant lines as  \Lya\ and \MgIIlambda. 
Additionally, we confirm some contribution from \OIlambda\ to the feature at 1280\,\AA, although there is also significant flux from \ion{Fe}{2} in the model. 
We also verify the identification of \CIIlambda\ with a small contribution from \ion{Si}{2} and absorption from \ion{Ni}{2}.
The observed \HeIIlambda\ emission is not present in the model spectra although the ion is in the model. 
As there is no feature due to another species in the model spectra at this location, we keep our \ion{He}{2} identification. 

We also identify previously unknown features using the models.
The majority of the flux at both epochs is due to \ion{Fe}{2}, and we identify the observational peaks at 1345\,\AA, 1470\,\AA, 1760\,\AA, 2575\,\AA, and 2920\,\AA\ as \ion{Fe}{2}. 
The peak at 1710\,\AA\ is a combination of \ion{N}{1} and \ion{Ni}{2} and the peak at 1790\,\AA\ is \ion{Si}{2}, although the strength of this feature is much stronger in the model than in the observation. 
If confirmed, this would be the first indication of nitrogen in the UV spectra.
At day 600, similar features are present, although at different strengths, and while the overall spectrum matches, there is a larger mismatch in the strength of individual features between the observation and the models. 

Examining the \Lya\ and \MgII\ line fluxes and shapes can tell us how well the model represents the optical depth in the ejecta and CSM. 
The model flux underestimates the \Lya\ flux at day 350 and overestimates it at day 600. 
Interestingly, the flux of \Lya\ is significantly higher when \ion{Fe}{2} is excluded, indicating that \ion{Fe}{2} is absorbing some of these photons. 
It is possible that this absorption is too strong at day 350. 
At day 600, the \Lya\ flux is overestimated.
The shape of \Lya\ is challenging to compare as the observed profile is significantly affected by ISM absorption.
At 350 days, the model reproduces the width of the \MgII\ emission well and only slightly overestimates the flux, indicating that the reduced velocity of the CDS in these models is a better representation of the physical conditions. 
However, the model profile has a flatter top,  indicating a lower optical depth than the observed line experiences.
These effects are even more pronounced at day 600.
This could be due to the simplicity of the implementation of the CDS density as a Gaussian profile with spherical symmetry in the 1D models. 
In reality, the CDS is expected to fragment \citep{1995Chevalier}.

\section{Discussion} \label{sec:discussion}
\subsection{The Origin of the Asymmetric Line Profiles}
The blueshifted, asymmetric line profiles are either due to an asymmetric CDS, an asymmetric CSM, or a symmetric CDS and CSM in which the far side of ejecta is obscured. 
We note that for a sample of objects, we expect the orientation of this asymmetry would be random rather than always on the far side of the ejecta. 
Our models represent the symmetric CSM case and the asymmetries in the profiles of both \Lya\ and \MgII\ (two very strong resonant scattering lines) are present in the models with and without dust. 
This indicates that even at late phases, the line asymmetry can be an effect of the optical depth of the line transition in the CDS and does not necessarily imply dust formation.
The evolution of the line profiles to be more symmetric with time is also more consistent with the CDS becoming optically thin with time as it expands and cools (and possibly fragments).
If the asymmetry were due to newly formed dust and more dust formed over time, the line asymmetry would become stronger at later phases.
With this interpretation, the more symmetric \MgII\ profile of SN~2024ggi indicates a lower optical depth in the CDS.
Given that the early, transient narrow features in SN~2024ggi disappeared much faster than in SN~2023ixf (which shows a smaller radial extent of the CSM), it is not surprising that the CDS, which is mostly formed from the high-density CSM responsible for the early narrow features, would be smaller and therefore have a lower optical depth.

\subsection{Mass-Loss History}
As the supernova ejecta expand into the surrounding CSM, the ejecta and CSM interact, producing observable signatures which can be used to map the mass-loss history of the progenitor star.
When the ejecta hit CSM, they create a forward shock which moves out into the CSM and a reverse shock which moves into the ejecta, and a CDS can form between the two shocks. 
\citet{2025Nayana} find that the forward shock in SN~2023ixf is adiabatic by day 10 and does not contribute to the optical and UV flux. 
If a CDS is present, the reverse shock will be radiative \citep{1996Fransson}.
Additionally, the CDS will thermalize X-rays from the reverse shock, emitting them as UV and optical radiation, and the X-ray luminosity will be dominated by the forward shock.
For a radiative, cooling reverse shock, the total emitted energy is given by Equation 3.17 of \citet{1996Fransson}, which, assuming a power-law wind density profile with index 2 (steady-state wind) and a power-law density profile for the outer stellar envelope with index 12 \citep[for RSGs;][]{2017Chevalier}, can be written as
\begin{equation} \label{eqn:revshock}
    L_{\mathrm{rev}} = 1.13\times10^{40}
    \frac{\dot{M}_{-5}}{v_{w1}}\times V_{4}^{3}~\mathrm{erg~s^{-1}}
\end{equation}
where $\dot{M}_{-5}$ is the mass-loss rate in units of $10^{-5}$ \masslossunit, $v_{w1}$ is in units of $10$ \kms, and $V_{4}$ is the ejecta velocity at the reverse shock in units of $10^4$ \kms.
Half of this energy will be radiated toward the CDS and half toward the ejecta. 
If the CDS reprocesses all of this energy into the UV and optical, then the shock power in the CMFGEN models is equivalent to half of the reverse-shock luminosity. 

\begin{figure*}
    \centering
    \includegraphics[width=1\textwidth]{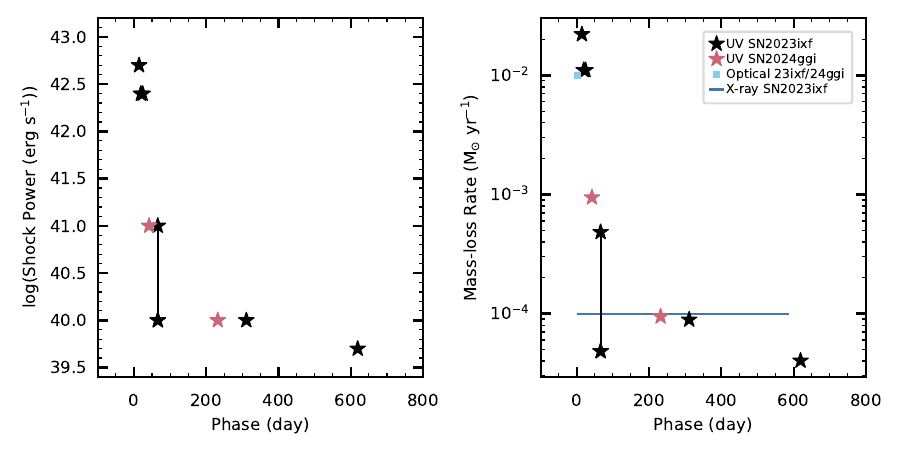}
    \caption{The shock luminosity {\it (left)} and mass-loss rate {\it (right)} for SN~2023ixf (black) and SN~2024ggi (pink) derived from UV observations (stars), X-ray observations (blue line; Jacobson-Galan et al. in prep), and narrow optical lines \citep[squares; ][]{2023Jacobson-Galan,2023Bostroem2, 2024Shrestha}. 
    The UV observation of SN~2023ixf at 66\,d fell between two models, so we display both here connected with a vertical line. }
    \label{fig:shock-density}
\end{figure*}

We use \autoref{eqn:revshock} to calculate the mass-loss rate of SN~2023ixf and SN~2024ggi from the shock-power luminosity of the CMFGEN model that best matches the overall UV flux of our observations. 
For the 15--66 day spectra of SN~2023ixf we adopt the shock power found by \citet{2024Bostroem}.
We use a wind velocity of 25 \kms\ \citep{2025Dickinson} for SN~2023ixf and 
37 \kms\ for SN~2024ggi \citep{2024Shrestha}. 
We used $V_{\mathrm{max},{\rm blue}}$ from the \MgII\ line in \autoref{tab:velocity}  as a proxy for the reverse-shock velocity as it measures the velocity of the CDS. 
For the velocity of the reverse shock at 66\,d, we measure the \MgII\ emission line in the same manner as described in this paper finding $V_{\mathrm{max,blue}}=9700$ \kms\ with an uncertainty of 440 \kms. 
We cannot measure $V_{\mathrm{max},{\rm blue}}$ from \MgII\ in earlier epochs as \MgII\ is a P Cygni profile.
For these epochs we assumed a reverse shock velocity of 10000\,\kms.
For SN~2024ggi, we use the $V_{\mathrm{max,blue}}$ of \MgII\ at 232\,d. 
The mass-loss rates we calculate are given in \autoref{tab:massloss} and shown in the right panel of \autoref{fig:shock-density}.
We note the mass-loss rate we find are not a steady-state wind, as assumed in the calculations, but leave more sophisticated modeling to future work.


To mass-loss history we derived from the UV luminosity of SN~2023ixf and SN~2024ggi we add the mass-loss rates inferred from the CMFGEN models of the early-time, narrow emission of SN~2023ixf and SN~2024ggi and the X-ray mass-loss rate of SN~2023ixf.
SN~2023ixf and SN~2024ggi showed narrow, high-ionization emission lines in the optical which faded over the first week. 
Optical observations were compared to the models of \citet{2017Dessart} and it was found that the best match was to the r1w4 and r1w6 models \citep{2023Bostroem2, 2023Jacobson-Galan, 2024Shrestha}.
The mass-loss rates are taken directly from \citet{2017Dessart}.
In the right panel of \autoref{fig:shock-density} we show the mass-loss rate of $\dot{M}=1\times10^{-4}$ \masslossunit\ that Jacobson-Galan et al. (in prep.) calculate from the X-ray emission measure for SN~2023ixf.

We calculate the expansion radius from our CDS velocity and use the wind velocity to determine how many years before explosion our calculated mass-loss rates occurred. 
With the observations in this paper we trace both profiles back to a radius of $r=4.6\times10^{16}$ cm or $\sim$600 yr before explosion for SN~2023ixf and $r=1.7\times10^{16}$ cm or $\sim$150 yr before explosion for SN~2024ggi.
The inferred mass-loss rate from the UV flux for SN~2023ixf decreases with time from $10^{-2}$ to $10^{-5}$ \masslossunit.
It starts off higher than the mass-loss rate calculated from the X-ray luminosity but is consistent with the mass-loss rate inferred from the transient, narrow emission in the early-time spectra.  
This represents a phase of enhanced mass loss. 
By day 199, the mass-loss rate is in the normal range for a quiescent RSG  \citep{2014SmithARAA} and is now below the X-ray mass-loss rate. 
The later-time discrepancy between the mass-loss rate inferred from the UV flux and the X-ray luminosity could be due to our assumption that the reverse shock is fully reprocessed by the CDS. 
If some X-ray flux were to escape, it would increase the X-ray flux and decrease the flux observed in the UV. 
This would then require a higher mass-loss rate to reproduce the UV flux and a lower mass-loss rate to reproduce the X-ray flux. 
While Jacobson-Galan et al. (in prep.) are not able to confirm the presence of the reverse shock, they note a flattening of the X-ray luminosity which could indicate X-ray flux from the reverse shock. 
Higher S/N X-ray observations are required to confirm or rule out this scenario.

We note that this calculation deviates significantly from that of \citet{2024Bostroem}, which considers the shock power from CMFGEN to be from the forward shock and conversion of CSM mass swept up by the forward shock into kinetic energy. 
\citet{2025Nayana} find that this description is not valid for SN~2023ixf for all epochs they model ($>4$ days) as the free-free timescale is greater than the dynamical timescale.
Our mass-loss rates are about an order of magnitude higher than the calculated rates in \citet{2024Bostroem} for day 66.

\begin{table*}[]
    \centering
    \caption{CSM properties for SN~2023ixf and SN~2024ggi. The shock power at days 14--66 is from \citet{2024Bostroem}\label{tab:massloss}}
    \begin{tabular}{|c|c|c|c|}
    \hline
    Phase & Time Before Explosion & Absorbed Shock Power & Mass-Loss Rate \\
    (day)& (year)&(\pwrunit) & (\masslossunit) \\
    \hline
    \multicolumn{4}{|c|}{SN~2023ixf}\\
    \hline
    14 & 14 & $5\times10^{42}$    & $2\times10^{-2}$ \\
    19 & 19 & $2.5\times10^{42}$  & $1\times10^{-2}$ \\
    24 & 24 & $2.5\times10^{42}$ &  $1\times10^{-2}$ \\
    66 & 65 & $1\times10^{40}$ -- $1\times10^{41}$   & $5\times10^{-5}$ -- $5\times10^{-4}$ \\
    311 & 306 & $1\times10^{40}$   & $9\times10^{-5}$ \\
    619 & 610 & $5\times10^{39}$   & $4\times10^{-5}$ \\
    \hline
    \multicolumn{4}{|c|}{SN~2024ggi}\\
    \hline
    41 & 28 & $1\times10^{41}$ & $1\times10^{-3}$\\
    232 & 155 & $1\times10^{40}$ &  $9\times10^{-5}$\\
    \hline
    \end{tabular}

\end{table*}

\subsection{Comparison to Other Supernovae}
\begin{figure*}
 \centering
     \includegraphics[width=\textwidth]{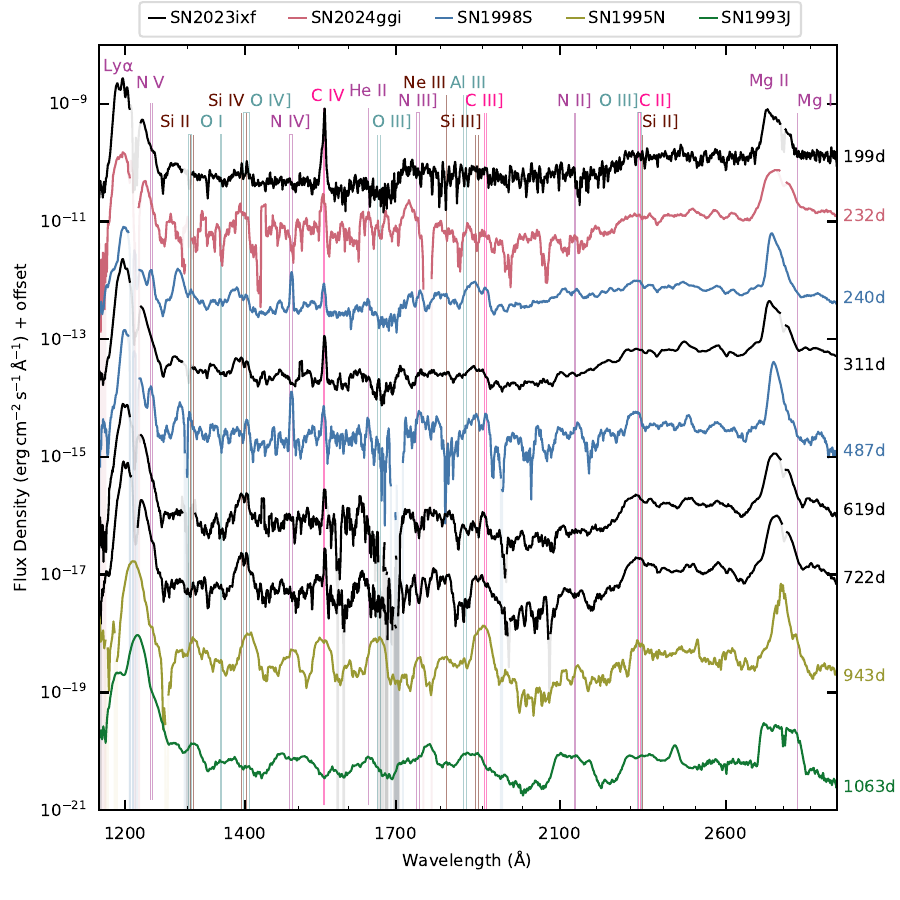}
     \caption{SN~2023ixf (black) and SN~2024ggi (pink) in the context of UV spectra observed with HST between $\sim$200\,d and 1000\,d. 
     SN~1998S (blue) is similar to SN~2023ixf and SN~2024ggi, although the \Lya\ and \MgII\ profiles have more red-side absorption and narrow nitrogen lines are present which are not in SN~2023ixf nor SN~2024ggi. SN~1993J (yellow) and SN~1995N (green) show two possible evolutions for SN~2023ixf and SN~2024ggi, with SN~1993J showing rounded profiles while SN~1995N shows pronounced boxy lines.}
     \label{fig:SNcompare}
 \end{figure*}

UV observations of Type II supernovae are sparse and generally limited to those that are strongly interacting (i.e., persistent narrow lines). 
Thus, it is challenging to compare our observations to other supernovae at the same epoch.
However, we can create a temporal series using all HST UV observations (SN~1993J, SN~1995N, and SN~1998S) which we present in \autoref{fig:SNcompare}.
Overall, there is surprising similarity for events whose initial evolution was different. 
At these late times, the UV (and most of the optical by day 619) emission is dominated by emission from the CDS at the CSM-ejecta interface. 
Given that all of these events showed early CSM interaction, they all are expected to have a CDS.

SN~1993J is a Type IIb supernova, initially showing weak H lines that faded over time and strong He lines  \citep{1993Filippenko}. 
It is thought that a Type IIb supernova is caused by a star that has lost most of its hydrogen envelope \citep{1993Filippenko, 1994Filippenko} and thus has significant CSM surrounding it.
SN~1993J showed signs of interaction in the early and late optical spectra and at other wavelengths throughout its evolution \citep{1994Benetti, 1994vanDyk,1994Fransson, 1994Zimmermann,1994Filippenko,1995Patat,2000Matheson, 2000Matheson2, 2017Smith2}.
The late-time UV evolution, from 670\,d to 2585\,d in the NUV and 1063--2585\,d in the FUV, was detailed by \citet{2005Fransson}. 
The first complete FUV+NUV spectrum (1063\,d) taken with the Faint Object Spectrograph (FOS) on HST is shown in \autoref{fig:SNcompare}.
For SN~1993J, we use an explosion epoch of 1993-03-27 12:00:00 \citep{1994Lewis}, Milky Way extinction $E(B-V)=0.069$\,mag \citep{2011Schlafly}, host extinction $E(B-V)=0.10\pm0.11$\,mag \citep{2014Ergon}, and a distance modulus of $\mu=27.80\pm0.10$\,mag \citep{1994Freedman}.

SN~1998S is a luminous Type II supernova which showed narrow optical features at early times that faded over the first few weeks of evolution, indicating a shell of confined, high-density CSM \citep{2001Fassia,2001Anupama, 2000Leonard,2000Liu,2001Lentz, 2015Shivvers}. 
Even after these lines faded, SN~1998S showed other signs of CSM interaction \citep[e.g.,][]{2002Pooley,2004Pozzo,2012Mauerhan,2017Smith2}. 
Before SN~2023ixf, SN~1998S had the most comprehensive UV dataset for this time period, with spectra from HST/STIS at 28\,d, 72\,d, 238\,d, and 485\,d \citep{2005Fransson}. 
For SN~1998S, we use an explosion epoch of 1998-02-27 04:33:36.000 UT \citep{2023Bostroem2}, Milky Way extinction $E(B-V)=0.0202\pm0.0009$\,mag \citep{2011Schlafly}, host extinction $E(B-V)=0.2\pm0.02$\,mag \citep{2000Leonard}, and distance modulus $\mu=31.18\pm0.38$\,mag \citep{1997Willick}.

SN~1995N is an Type IIn supernova which was discovered $\sim$10 months after explosion with persistent CSM interaction (narrow emission lines) in the optical. 
An HST/FOS UV spectrum was obtained on day 943 and presented by \citet{2002Fransson}.
For SN~1995N, we adopt an explosion epoch of 1994-07-04 00:00:00 UT \citep{2002Fransson}, total extinction $E(B-V)=0.11$\,mag \citep{2002Fransson}, and distance $d=24$ Mpc \citep{2002Fransson}.

\citet{2005Fransson} note that in SN~1993J, the \MgIIlambda\ profile is asymmetric at day 670 with a stronger blue side and blueshifted.
This asymmetry decreases with time such that by day 1399 it is symmetric.
Compared to SN~1993J, the \MgII\ profile in SN~2023ixf is less boxy and more asymmetric. 
However, over time it evolves to be more symmetric, and it will be interesting to see if future observations show if it evolves into a boxy profile indicating emission from a shell. 
Combining the day 1063 and day 1399 spectra and fitting the \MgII\ and \NII\  lines, \citet{2005Fransson} find an outer velocity of 10,000 \kms, a little higher than that found for SN~2023ixf and SN~2024ggi.

The \Lya\ and \MgII\ profiles of SN~1998S showed strong asymmetry at day 72 which persists throughout the evolution.
Like SN~2023ixf and SN~2024ggi, they find that the \Lya\ and \MgII\ profiles are very similar at day 72 and later epochs.
The velocity of the blue edge of the \MgII\ line at day 72 is 8000 \kms\ \citep{2005Fransson}, slightly decreasing with time to $\sim$7000 \kms.
The velocity of the blue edge of the UV lines indicates the maximum velocity of the CDS. 
As the CDS is formed when the supernova ejecta run into the CSM, the speed of the CDS is a product of the ejecta velocity and the density of the CSM,  which will slow down at the ejecta interface. 
Thus, the higher speed of this edge in SN~2023ixf and SN~2024ggi indicates that the density of material surrounding SN~1998S was greater or that the ejecta velocity was lower. 
Given the persistence of the narrow features in SN~1998S and the high luminosity, we prefer CSM density as the dominant source of the lower velocity in SN~1998S.


Nitrogen has been detected in SN~1979C, SN~1987A, SN~1993J, SN~1995N, and SN~1998S, in the form of narrow lines (indicating CSM origin) in SN~1987A and SN~1998S and in broad lines (indicating ejecta origin) 
for the remainder. 
These features are significantly weaker, if present at all, in the spectra of SN~2023ixf and SN~2024ggi.
Nitrogen enhancement is a product of the CNO cycle as $^{14}\mathrm{N}(p,\gamma)^{15}\mathrm{O}$ is the slowest process.
 Nitrogen enrichment has been qualitatively used to indicate substantial mass loss and in some cases extrapolated to more massive progenitors \citep{2002Fransson, 2005Fransson, 2019Boian}.
Nitrogen enrichment is determined from the intensity ratios of \CIIIlambda, 
\CIVlambda, \NIIIlambda, \NIVlambda, and \OIIIlambda\ based on the assumption that \CIII\ and \NIII\ form in the same region, \CIV\ and \NIV\ form in the same region, and \NIII\ and \NIV\ form in the same region as \OIII\ \citep{1982Kallman}.  

Although narrow \CIV\ is observed in both SN~2023ixf and SN~2024ggi, narrow \NIV, \CIII, \NIII, or \OIII\ are not detected in the spectra at any epoch. 
We also do not detect broad \NIV, \NIII, \CIV\, or \OIII.
Interestingly, narrow, high-ionization optical lines present in the first week after explosion did reveal \NV, \NIV, and \NIII.
It is possible that nitrogen is present at lower ionization and we tentatively detect \ion{N}{1} based on the models. 
Further modeling is needed to definitively interpret the lack of highly ionized nitrogen in the late-time UV spectra of SN~2023ixf and SN~2024ggi. 

SN~1993J and SN~1995N illustrate what the future evolution of SN~2023ixf and SN~2024ggi might look like. 
While the UV line profiles of SN~1993J are flat-topped, the broad profiles of SN~1995N are more rounded, similar to those of SN~2023ixf. 
\citet{2002Fransson} suggest that aspherical CSM where the highest density material is equatorial could explain these rounded profiles.
This is consistent with polarization measurements of SN~2023ixf in \citet{2023Vasylyev2}, \citet{2024Singh}, and \citet{2025Shrestha}, and with high-resolution optical spectroscopy of SN~2023ixf \citep{2023Smith, 2025Dickinson}, which suggest an equatorial torus of dense CSM.

\section{Summary} 
\label{sec:summary}
We present FUV and NUV spectra of SN~2023ixf at average phases of 199, 311, 619, and 722\,d and SN~2024ggi at days 41 (NUV only) and 232. 
These are the first FUV spectra presented for both supernovae. 
Most UV lines exhibit broad (7000--9000 \kms), asymmetric, and blueshifted emission 
 and we identify narrow, unresolved emission in \CIVlambda. 
While the broad lines originate in the CDS, the narrow lines are too narrow to form in material that has been accelerated by the ejecta and therefore likely originate from the surrounding CSM.

Over time, the blue-side maximum velocity of the broad lines decreases and the red-side maximum velocity increases.
At the same time, the line optical depth decreases, resulting in more symmetric (but still blueshifted) profiles. 
We find that the UV line profiles are very similar and compare them to the optical emission lines. 
At day 200, the optical profiles are distinct from the UV profiles, with maximum velocities of 3000 \kms.
This is consistent with the UV emission originating in the outer regions of the CDS powered by ejecta-CSM interaction while the optical lines originate in the inner ejecta powered by radioactive decay. 
By day 619, the optical and UV profiles are virtually the same, indicating that the power from CSM emission is dominating over the radioactive-decay power in the optical. 

We model the spectra of SN~2023ixf and SN~2024ggi with \texttt{CMFGEN} using the models of \citet{2022Dessart} for SN~2024ggi and building custom models for SN~2023ixf at days 300 and 600, which include both absorbed shock power, scaling of the radioactive decay power, and silicate dust. 
We find that some absorbed shock power is required to reproduce the spectra at all epochs. 
At day 42, we find that the UV+optical spectrum of SN~2024ggi is best matched by a model with shock power from CSM interaction $P_{\mathrm{shock,abs}}=1\times10^{41}$ \pwrunit. 
At day 232, the shock power has decreased to $P_{\mathrm{shock,abs}}=1\times10^{40}$ \pwrunit.
For SN~2023ixf, we find  $P_{\mathrm{shock,abs}}=1\times10^{40}$ \pwrunit\ at day 311 and  $P_{\mathrm{shock,abs}}=5\times10^{39}$ \pwrunit\ at day 619. 
We note that dust is not required to produce the asymmetric shape of the UV lines, indicating that this profile is not a definitive indicator dust formation.
We decompose the model spectra into individual species, confirming many of our earlier line identifications. 
Additionally, we identify \ion{Fe}{2} features as well as a possible \NIlambda\ line around 1710\,\AA. 

We use the absorbed shock power determined from the models to derive the mass-loss history of SN~2023ixf and SN~2024ggi. 
For SN~2023ixf, we find mass-loss rates decrease from $2\times10^{-2}$ \masslossunit\ to $\sim5\times10^{-4}$ \masslossunit\ over the first 15--60 days and $9\times10^{-5}$ \masslossunit\ at day 311 and and $4\times10^{-5}$ \masslossunit\ at day 619. 
A lower mass-loss rate of $1\times10^{-3}$ \masslossunit\ is obtained for the early day 41 epoch of SN~2024ggi which decreases to $9\times10^{-5}$ \masslossunit\ at day 232. 
We compare the mass-loss rate derived from X-ray observations and find that while the UV mass-loss rate decreases with time, the X-ray mass-loss rate remains constant.
These mass-loss rates correspond to 30--150 yr before explosion for SN~2024ggi and 15--600 yr before explosion for SN~2023ixf.

Finally, we place SN~2024ggi and SN~2023ixf in the context of other Type II supernovae with late-time UV observations from HST. 
In all spectra the dominant lines are \Lya\ and \MgII\ although the line profiles vary, indicating differences in the optical depth of the CDS of each supernova. 
Narrow \CIV, which is present in both SN~2023ixf and SN~2024ggi, is clearly visible in SN~1998S, and broad \CIV\ is detected in SN~1995N. 
While all three comparison supernovae had highly ionized nitrogen lines, we do not find these in SN~2023ixf or SN~2024ggi, which could indicate a lack of nitrogen enrichment or that these ionization levels are not populated. 

With the spectra of SN~2023ixf and SN~2024ggi filling in gaps at $\sim300$, 600, and 700\,d, 
we now have temporal sampling of $\sim100$\,d for the first 1000\,d of evolution and can compare spectra across multiple objects at day 200. 
These observations are the only UV spectra of Type II supernovae at this phase since HST's launch in 1990. 
In the local Universe, with new UV telescopes on the horizon such as Ultrasat \citep{2024Shvartzvald} and UVEX \citep{2021Kulkarni}, these spectra help ensure that instrument designs are well suited to the relevant scientific questions. 
Additionally, as JWST, the Rubin Observatory's Legacy Survey of Space and Time, and the Roman Space Telescope High-latitude Time Domain Survey are pushing supernova observations to high redshift, often with low-S/N spectra or only photometry, it is crucial to understand the UV properties of Type II supernovae to interpret these observations.

\section{acknowledgments}
K.A.B. is supported by an LSST-DA Catalyst Fellowship; this publication was thus made possible through the support of grant 62192 from the John Templeton Foundation to LSST-DA.
Time-domain research by the UC Davis team
and S.V. is supported by U.S. National Science
Foundation (NSF) grant AST-2407565.
Time-domain research by the University of Arizona team
and D.J.S. is supported by
NSF grants 2108032, 2308181, 2407566, and 2432036,
and by the Heising-Simons Foundation under grant \#2020-1864. 
Supernova research at Rutgers University is supported by NSF award AST-2407567.
The LCO group is supported by NSF grants AST-1911151 and AST-2308113.
N.F. acknowledges support from the NSF Graduate Research Fellowship Program under grant  DGE-2137419.
A.V.F. is grateful for financial support
from NASA/HST grants GO-16178 and GO-16656 from the Space Telescope Science Institute 
(STScI), which is operated by the Association of Universities for Research in Astronomy (AURA), Inc., under NASA contract NAS5-26555; additional support was provided by individual donors. 

This work makes use of observations from the Las Cumbres Observatory network.
Based in part on observations obtained at the international Gemini Observatory (GN-2025A-Q-20; PI K. Azalee Bostroem), a program of NSF's NOIRLab, which is managed by AURA, Inc., under a cooperative agreement with the NS. On behalf of the Gemini Observatory partnership: the National Science Foundation (United States), National Research Council (Canada), Agencia Nacional de Investigaci\'{o}n y Desarrollo (Chile), Ministerio de Ciencia, Tecnolog\'{i}a e Innovaci\'{o}n (Argentina), Minist\'{e}rio da Ci\^{e}ncia, Tecnologia, Inova\c{c}\={o}es e Comunica\c{c}\={o}es (Brazil), and Korea Astronomy and Space Science Institute (Republic of Korea).

The Legacy Surveys consist of three individual and complementary projects: the Dark Energy Camera Legacy Survey (DECaLS; Prop. ID 2014B-0404, PIs David Schlegel and Arjun Dey), the Beijing-Arizona Sky Survey (BASS; NOAO Prop. ID 2015A-0801, PIs Zhou Xu and Xiaohui Fan), and the Mayall z-band Legacy Survey (MzLS; Prop. ID 2016A-0453, PI Arjun Dey). 
DECaLS, BASS, and MzLS together include data obtained, respectively, at the Blanco telescope, Cerro Tololo Inter-American Observatory, NSF's NOIRLab; the Bok telescope, Steward Observatory, University of Arizona; and the Mayall telescope, Kitt Peak National Observatory, NOIRLab. 
Pipeline processing and analyses of the data were supported by NOIRLab and the Lawrence Berkeley National Laboratory (LBNL). 
The Legacy Surveys project is honored to be permitted to conduct astronomical research on Iolkam Du'ag (Kitt Peak), a mountain with particular significance to the Tohono O'odham Nation.
NOIRLab is operated by AURA, Inc., under a cooperative agreement with the NSF.
LBNL is managed by the Regents of the University of California under contract to the U.S. Department of Energy.

This project used data obtained with the Dark Energy Camera (DECam), which was constructed by the Dark Energy Survey (DES) collaboration. Funding for the DES Projects has been provided by the U.S. Department of Energy, the NSF, the Ministry of Science and Education of Spain, the Science and Technology Facilities Council of the United Kingdom, the Higher Education Funding Council for England, the National Center for Supercomputing Applications at the University of Illinois at Urbana-Champaign, the Kavli Institute of Cosmological Physics at the University of Chicago, the Center for Cosmology and Astro-Particle Physics at the Ohio State University, the Mitchell Institute for Fundamental Physics and Astronomy at Texas A\&M University, Financiadora de Estudos e Projetos, Fundacao Carlos Chagas Filho de Amparo, Financiadora de Estudos e Projetos, Fundacao Carlos Chagas Filho de Amparo a Pesquisa do Estado do Rio de Janeiro, Conselho Nacional de Desenvolvimento Cientifico e Tecnologico and the Ministerio da Ciencia, Tecnologia e Inovacao, the Deutsche Forschungsgemeinschaft, and the Collaborating Institutions in the Dark Energy Survey. The Collaborating Institutions are Argonne National Laboratory, the University of California at Santa Cruz, the University of Cambridge, Centro de Investigaciones Energeticas, Medioambientales y Tecnologicas-Madrid, the University of Chicago, University College London, the DES-Brazil Consortium, the University of Edinburgh, the Eidgenossische Technische Hochschule (ETH) Zurich, Fermi National Accelerator Laboratory, the University of Illinois at Urbana-Champaign, the Institut de Ciencies de l'Espai (IEEC/CSIC), the Institut de Fisica d'Altes Energies, Lawrence Berkeley National Laboratory, the Ludwig Maximilians Universitat Munchen and the associated Excellence Cluster Universe, the University of Michigan, NSF's NOIRLab, the University of Nottingham, the Ohio State University, the University of Pennsylvania, the University of Portsmouth, SLAC National Accelerator Laboratory, Stanford University, the University of Sussex, and Texas A\&M University.

BASS is a key project of the Telescope Access Program (TAP), which has been funded by the National Astronomical Observatories of China, the Chinese Academy of Sciences (the Strategic Priority Research Program ``The Emergence of Cosmological Structure,'' grant  XDB09000000), and the Special Fund for Astronomy from the Ministry of Finance. BASS is also supported by the External Cooperation Program of Chinese Academy of Sciences (grant  114A11KYSB20160057) and the Chinese National Natural Science Foundation (grant  12120101003,  11433005).

The Legacy Survey team makes use of data products from the Near-Earth Object Wide-field Infrared Survey Explorer (NEOWISE), which is a project of the Jet Propulsion Laboratory/California Institute of Technology. NEOWISE is funded by the National Aeronautics and Space Administration.
The Legacy Surveys imaging of the DESI footprint is supported by the Director, Office of Science, Office of High Energy Physics of the U.S. Department of Energy under Contract DE-AC02-05CH1123; by the National Energy Research Scientific Computing Center, a DOE Office of Science User Facility under the same contract; and by the NSF, Division of Astronomical Sciences under Contract AST-0950945 to NOAO.

\vspace{5mm}
\facilities{Gemini:North(GMOS), HST(STIS), HST(COS), LCOGT, MAST (HASP), MMT(Binospec)}


\software{astropy \citep{astropy_collaboration_astropy_2013, astropy_collaboration_astropy_2018, astropy_collaboration_astropy_2022}, BANZAI \citep{2018McCully}, Binospec \citep{2019Kansky}, DRAGONS \citep{2019Labrie}, Hubble Advanced Spectral Products \citep{2024Debes}, IRAF \citep{1986Tody, 1993Tody, 2024Fitzpatrick}, lcogtsnpipe \citep{2016Valenti}, Light Curve Fitting \citep{LCFitting}, MatPLOTLIB \citep{hunter_matplotlib_2007}, NumPy \citep{harris_array_2020},  SciPy \citep{2020Virtanen}, specutils \citep{2025Earl}, stistools \citep{2019Sohn} WISeREP \citep{2017Yaron}  } 



\appendix
\counterwithin{figure}{section}
\section{Line Profiles for Custom Models \label{sec:appendix}}
\begin{figure}
    \centering
    \includegraphics[width=0.9\textwidth]{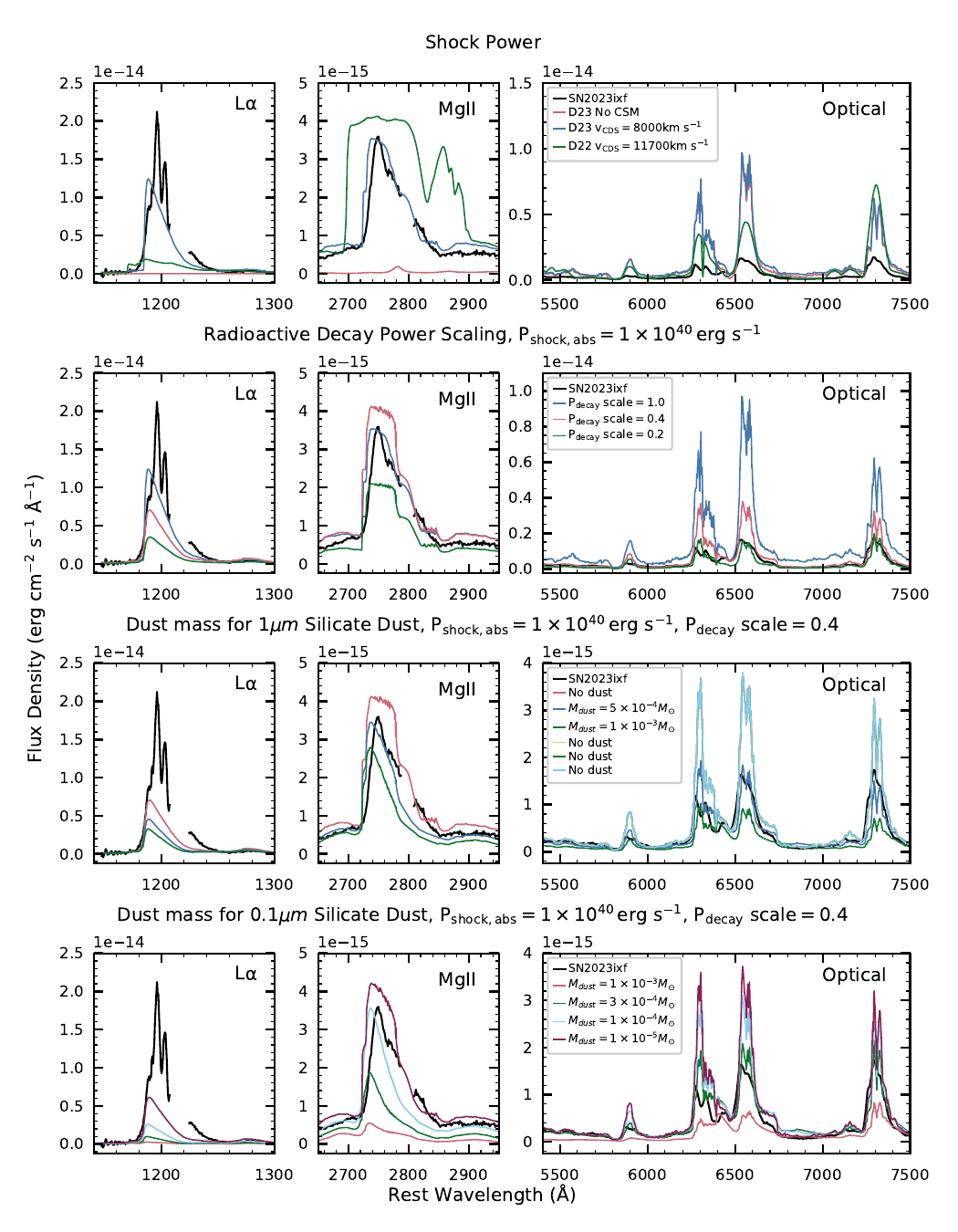}
    \caption{The effects of absorbed shock power (from CSM interaction; top row), scaling of radioactive decay power (second row from the top), 1\micron\ silicate dust (third row from the top), and 0.1\micron\ silicate dust (bottom) on \Lya\ (left column), \MgII\ (middle column), and optical (centered on \Ha; right column). \Lya\ and \MgII are only present in the models with CSM interaction. Both scaling the radioactive decay and dust have the largest effect in the optical. }
    \label{fig:ModelLineFlux}
\end{figure}

\autoref{fig:ModelLineFlux} shows the models in \autoref{fig:all_model} in linear flux space focusing on \Lya\ (left column), \MgIIlambda\ (middle column), and the optical (right column) including [\ion{O}{1}], \Ha\, and [\ion{Ca}{2}]. The effects of scaling the radioactive power absorption and dust on the individual line profiles can be seen more easily, especially in the optical. 

\bibliography{references}{}
\bibliographystyle{aasjournal}
\end{document}

%% file: new-commands.tex
\newcommand{\msun}{$M_\mathrm{\odot}$}

\newcommand{\Ha}{$\mathrm{H\alpha}$}
\newcommand{\Lya}{$\mathrm{Ly\alpha}$}

\newcommand{\OIlambda}{\ion{O}{1} $\lambda\lambda$1302.5--1304.9}
\newcommand{\OI}{\ion{O}{1}}
\newcommand{\OIIIlambda}{\ion{O}{3}] $\lambda\lambda1660.8,1666.1$}
\newcommand{\OIII}{\ion{O}{3}]}
\newcommand{\MgIIlambda}{\ion{Mg}{2} $\lambda\lambda$2795.5,2802.7}
\newcommand{\MgII}{\ion{Mg}{2}}
\newcommand{\CIV}{\ion{C}{4}}
\newcommand{\CIVlambda}{\ion{C}{4} $\lambda\lambda1548.9,1550.8$}
\newcommand{\SiIV}{\ion{Si}{4}}
\newcommand{\SiIVlambda}{\ion{Si}{4} $\lambda\lambda1393.8,1402.8$}

\newcommand{\HeIIlambda}{\ion{He}{2} $\lambda1640$}
\newcommand{\SiIII}{\ion{Si}{3}]}
\newcommand{\SiIIIlambda}{\ion{Si}{3}] $\lambda\lambda1882.7,1892.0$}
\newcommand{\CIII}{\ion{C}{3}]}
\newcommand{\CIIIlambda}{\ion{C}{3}] $\lambda\lambda1906.7,1909.6$}
\newcommand{\CII}{\ion{C}{2}]}
\newcommand{\CIIlambda}{\ion{C}{2}] $\lambda\lambda$2323.5--2328.1}

\newcommand{\NV}{\ion{N}{5}]}
\newcommand{\NVlambda}{\ion{N}{5}] $\lambda\lambda1238.8,1242.8$}
\newcommand{\NIV}{\ion{N}{4}]}
\newcommand{\NIVlambda}{\ion{N}{4}] $\lambda\lambda$1483.3--1487.9}
\newcommand{\NIII}{\ion{N}{3}]}
\newcommand{\NIIIlambda}{\ion{N}{3}]] $\lambda\lambda$1746.8--1754.0}
\newcommand{\NII}{\ion{N}{2}]}

\newcommand{\NIlambda}{\ion{N}{1} $\lambda\lambda 1742.7,1745.3$}

\newcommand{\kms}{$\mathrm{km~s^{-1}}$}
\newcommand{\pwrunit}{$\mathrm{erg~s^{-1}}$}

\newcommand{\masslossunit}{$M\mathrm{_{\odot}~yr^{-1}}$}

%% file: affiliations.tex
\newcommand{\LCO}{\affiliation{Las Cumbres Observatory, 6740 Cortona Drive, Suite 102, Goleta, CA 93117-5575, USA}}
\newcommand{\UCSB}{\affiliation{Department of Physics, University of California, Santa Barbara, CA 93106-9530, USA}}
\newcommand{\UCSD}{\affiliation{Department of Astronomy \& Astrophysics, University of California, San Diego, 9500 Gilman Drive, MC 0424, La Jolla, CA 92093-0424, USA}}
\newcommand{\KITP}{\affiliation{Kavli Institute for Theoretical Physics, University of California, Santa Barbara, CA 93106-4030, USA}}
\newcommand{\UCD}{\affiliation{Department of Physics and Astronomy, University of California, Davis, 1 Shields Avenue, Davis, CA 95616-5270, USA}}
\newcommand{\WIS}{\affiliation{Department of Particle Physics and Astrophysics, Weizmann Institute of Science, 76100 Rehovot, Israel}}
\newcommand{\OKC}{\affiliation{Oskar Klein Centre, Department of Astronomy, Stockholm University, Albanova University Centre, SE-106 91 Stockholm, Sweden}}
\newcommand{\OAPD}{\affiliation{INAF-Osservatorio Astronomico di Padova, Vicolo dell'Osservatorio 5, I-35122 Padova, Italy}}
\newcommand{\OAB}{\affiliation{INAF-Osservatorio Astronomico di Brera, Via E. Bianchi 46, I-23807, Merate (LC), Italy}}
\newcommand{\Caltech}{\affiliation{Cahill Center for Astronomy and Astrophysics, California Institute of Technology, Mail Code 249-17, Pasadena, CA 91125, USA}}
\newcommand{\GSFC}{\affiliation{Astrophysics Science Division, NASA Goddard Space Flight Center, Mail Code 661, Greenbelt, MD 20771, USA}}
\newcommand{\UMD}{\affiliation{Joint Space-Science Institute, University of Maryland, College Park, MD 20742, USA}}
\newcommand{\UCB}{\affiliation{Department of Astronomy, University of California, Berkeley, CA 94720-3411, USA}}
\newcommand{\TTU}{\affiliation{Department of Physics, Texas Tech University, Box 41051, Lubbock, TX 79409-1051, USA}}
\newcommand{\STScI}{\affiliation{Space Telescope Science Institute, 3700 San Martin Drive, Baltimore, MD 21218-2410, USA}}
\newcommand{\UT}{\affiliation{University of Texas at Austin, 1 University Station C1400, Austin, TX 78712-0259, USA}}
\newcommand{\IoA}{\affiliation{Institute of Astronomy, University of Cambridge, Madingley Road, Cambridge CB3 0HA, UK}}
\newcommand{\QUB}{\affiliation{Astrophysics Research Centre, School of Mathematics and Physics, Queen's University Belfast, Belfast BT7 1NN, UK}}
\newcommand{\IPAC}{\affiliation{IPAC, Mail Code 100-22, Caltech, 1200 E.\ California Blvd., Pasadena, CA 91125}}
\newcommand{\JPL}{\affiliation{Jet Propulsion Laboratory, California Institute of Technology, 4800 Oak Grove Dr, Pasadena, CA 91109, USA}}
\newcommand{\Southampton}{\affiliation{Department of Physics and Astronomy, University of Southampton, Southampton SO17 1BJ, UK}}
\newcommand{\LANL}{\affiliation{Space and Remote Sensing, MS B244, Los Alamos National Laboratory, Los Alamos, NM 87545, USA}}
\newcommand{\Tsinghua}{\affiliation{Physics Department and Tsinghua Center for Astrophysics, Tsinghua University, Beijing, 100084, People's Republic of China}}
\newcommand{\NAOC}{\affiliation{National Astronomical Observatory of China, Chinese Academy of Sciences, Beijing, 100012, People's Republic of China}}
\newcommand{\Itagaki}{\affiliation{Itagaki Astronomical Observatory, Yamagata 990-2492, Japan}}
\newcommand{\Einstein}{\altaffiliation{Einstein Fellow}}
\newcommand{\Hubble}{\altaffiliation{Hubble Fellow}}
\newcommand{\CfA}{\affiliation{Center for Astrophysics \textbar{} Harvard \& Smithsonian, 60 Garden Street, Cambridge, MA 02138-1516, USA}}
\newcommand{\UA}{\affiliation{Steward Observatory, University of Arizona, 933 North Cherry Avenue, Tucson, AZ 85721-0065, USA}}
\newcommand{\MPIA}{\affiliation{Max-Planck-Institut f\"ur Astrophysik, Karl-Schwarzschild-Stra\ss{}e 1, D-85748 Garching, Germany}}
\newcommand{\DSFP}{\altaffiliation{LSSTC Data Science Fellow}}
\newcommand{\HCO}{\affiliation{Harvard College Observatory, 60 Garden Street, Cambridge, MA 02138-1516, USA}}
\newcommand{\Carnegie}{\affiliation{Observatories of the Carnegie Institute for Science, 813 Santa Barbara Street, Pasadena, CA 91101-1232, USA}}
\newcommand{\TAU}{\affiliation{School of Physics and Astronomy, Tel Aviv University, Tel Aviv 69978, Israel}}
\newcommand{\Edinburgh}{\affiliation{Institute for Astronomy, University of Edinburgh, Royal Observatory, Blackford Hill EH9 3HJ, UK}}
\newcommand{\Birmingham}{\affiliation{Birmingham Institute for Gravitational Wave Astronomy and School of Physics and Astronomy, University of Birmingham, Birmingham B15 2TT, UK}}
\newcommand{\Bath}{\affiliation{Department of Physics, University of Bath, Claverton Down, Bath BA2 7AY, UK}}
\newcommand{\CTIO}{\affiliation{Cerro Tololo Inter-American Observatory, National Optical Astronomy Observatory, Casilla 603, La Serena, Chile}}
\newcommand{\Potsdam}{\affiliation{Leibniz-Institut f\"ur Astrophysik Potsdam (AIP), An der Sternwarte 16, D-14482 Potsdam, Germany}}
\newcommand{\INPE}{\affiliation{Instituto Nacional de Pesquisas Espaciais, Avenida dos Astronautas 1758, 12227-010, S\~ao Jos\'e dos Campos -- SP, Brazil}}
\newcommand{\UNC}{\affiliation{Department of Physics and Astronomy, University of North Carolina, 120 East Cameron Avenue, Chapel Hill, NC 27599, USA}}
\newcommand{\Ohio}{\affiliation{Astrophysical Institute, Department of Physics and Astronomy, 251B Clippinger Lab, Ohio University, Athens, OH 45701-2942, USA}}
\newcommand{\AAS}{\affiliation{American Astronomical Society, 1667 K~Street NW, Suite 800, Washington, DC 20006-1681, USA}}
\newcommand{\MMT}{\affiliation{MMT and Steward Observatories, University of Arizona, 933 North Cherry Avenue, Tucson, AZ 85721-0065, USA}}
\newcommand{\Geneva}{\affiliation{ISDC, Department of Astronomy, University of Geneva, Chemin d'\'Ecogia, 16 CH-1290 Versoix, Switzerland}}
\newcommand{\IUCAA}{\affiliation{Inter-University Center for Astronomy and Astrophysics, Post Bag 4, Ganeshkhind, Pune, Maharashtra 411007, India}}
\newcommand{\CMU}{\affiliation{Department of Physics, Carnegie Mellon University, 5000 Forbes Avenue, Pittsburgh, PA 15213-3815, USA}}
\newcommand{\NAOJ}{\affiliation{Division of Science, National Astronomical Observatory of Japan, 2-21-1 Osawa, Mitaka, Tokyo 181-8588, Japan}}
\newcommand{\IfA}{\affiliation{Institute for Astronomy, University of Hawai`i, 2680 Woodlawn Drive, Honolulu, HI 96822-1839, USA}}
\newcommand{\UCSC}{\affiliation{Department of Astronomy and Astrophysics, University of California, Santa Cruz, CA 95064-1077, USA}}
\newcommand{\Purdue}{\affiliation{Department of Physics and Astronomy, Purdue University, 525 Northwestern Avenue, West Lafayette, IN 47907-2036, USA}}
\newcommand{\Princeton}{\affiliation{Department of Astrophysical Sciences, Princeton University, 4 Ivy Lane, Princeton, NJ 08540-7219, USA}}
\newcommand{\Moore}{\affiliation{Gordon and Betty Moore Foundation, 1661 Page Mill Road, Palo Alto, CA 94304-1209, USA}}
\newcommand{\Durham}{\affiliation{Department of Physics, Durham University, South Road, Durham, DH1 3LE, UK}}
\newcommand{\JHU}{\affiliation{Department of Physics and Astronomy, The Johns Hopkins University, 3400 North Charles Street, Baltimore, MD 21218, USA}}
\newcommand{\Toronto}{\affiliation{David A.\ Dunlap Department of Astronomy and Astrophysics, University of Toronto,\\ 50 St.\ George Street, Toronto, Ontario, M5S 3H4 Canada}}
\newcommand{\Duke}{\affiliation{Department of Physics, Duke University, Campus Box 90305, Durham, NC 27708, USA}}
\newcommand{\NCU}{\affiliation{Graduate Institute of Astronomy, National Central University, 300 Jhongda Road, 32001 Jhongli, Taiwan}}
\newcommand{\Columbia}{\affiliation{Department of Physics and Columbia Astrophysics Laboratory, Columbia University, Pupin Hall, New York, NY 10027, USA}}
\newcommand{\Flatiron}{\affiliation{Center for Computational Astrophysics, Flatiron Institute, 162 5th Avenue, New York, NY 10010-5902, USA}}
\newcommand{\CIERA}{\affiliation{Center for Interdisciplinary Exploration and Research in Astrophysics, \\Northwestern University, 1800 Sherman Avenue, 8th Floor, Evanston, IL 60201, USA}}
\newcommand{\GeminiNorth}{\affiliation{Gemini Observatory, 670 North A`ohoku Place, Hilo, HI 96720-2700, USA}}
\newcommand{\Keck}{\affiliation{W.~M.~Keck Observatory, 65-1120 M\=amalahoa Highway, Kamuela, HI 96743-8431, USA}}
\newcommand{\UW}{\affiliation{Department of Astronomy, University of Washington, 3910 15th Avenue NE, Seattle, WA 98195-0002, USA}}
\newcommand{\Catalyst}{\altaffiliation{LSST-DA Catalyst Fellow}}
\newcommand{\USask}{\affiliation{Department of Physics and Engineering Physics, University of Saskatchewan, 116 Science Place, Saskatoon, SK S7N 5E2, Canada}}
\newcommand{\Thacher}{\affiliation{Thacher School, 5025 Thacher Road, Ojai, CA 93023-8304, USA}}
\newcommand{\Rutgers}{\affiliation{Department of Physics and Astronomy, Rutgers, the State University of New Jersey,\\136 Frelinghuysen Road, Piscataway, NJ 08854-8019, USA}}
\newcommand{\FSU}{\affiliation{Department of Physics, Florida State University, 77 Chieftan Way, Tallahassee, FL 32306-4350, USA}}
\newcommand{\Melbourne}{\affiliation{School of Physics, The University of Melbourne, Parkville, VIC 3010, Australia}}
\newcommand{\ASTROthreeD}{\affiliation{ARC Centre of Excellence for All Sky Astrophysics in 3 Dimensions (ASTRO 3D)}}
\newcommand{\Stromlo}{\affiliation{Mt.\ Stromlo Observatory, The Research School of Astronomy and Astrophysics, Australian National University, ACT 2601, Australia}}
\newcommand{\NCPAS}{\affiliation{National Centre for the Public Awareness of Science, Australian National University, Canberra, ACT 2611, Australia}}
\newcommand{\TAMU}{\affiliation{Department of Physics and Astronomy, Texas A\&M University, 4242 TAMU, College Station, TX 77843, USA}}
\newcommand{\Mitchell}{\affiliation{George P.\ and Cynthia Woods Mitchell Institute for Fundamental Physics \& Astronomy, College Station, TX 77843, USA}}
\newcommand{\ESO}{\affiliation{European Southern Observatory, Alonso de C\'ordova 3107, Casilla 19, Santiago, Chile}}
\newcommand{\ICE}{\affiliation{Institute of Space Sciences (ICE, CSIC), Campus UAB, Carrer
de Can Magrans, s/n, E-08193 Barcelona, Spain}}
\newcommand{\IEEC}{\affiliation{Institut d'Estudis Espacials de Catalunya, Gran Capit\`a, 2-4, Edifici Nexus, Desp.\ 201, E-08034 Barcelona, Spain}}
\newcommand{\Warwick}{\affiliation{Department of Physics, University of Warwick, Gibbet Hill Road, Coventry CV4 7AL, UK}}
\newcommand{\Macquarie}{\affiliation{School of Mathematical and Physical Sciences, Macquarie University, NSW 2109, Australia}}
\newcommand{\AAARC}{\affiliation{Astronomy, Astrophysics and Astrophotonics Research Centre, Macquarie University, Sydney, NSW 2109, Australia}}
\newcommand{\Capodimonte}{\affiliation{INAF - Capodimonte Astronomical Observatory, Salita Moiariello 16, I-80131 Napoli, Italy}}
\newcommand{\INFNNapoli}{\affiliation{INFN - Napoli, Strada Comunale Cinthia, I-80126 Napoli, Italy}}
\newcommand{\ICRANet}{\affiliation{ICRANet, Piazza della Repubblica 10, I-65122 Pescara, Italy}}
\newcommand{\MSU}{\affiliation{Center for Data Intensive and Time Domain Astronomy, Department of Physics and Astronomy,\\Michigan State University, East Lansing, MI 48824, USA}}
\newcommand{\IAP}{\affiliation{Institut d'Astrophysique de Paris, CNRS-Sorbonne Universit\'e, 98 bis boulevard Arago, 75014 Paris, France}}
\newcommand{\Pitt}{\affiliation{Department of Physics and Astronomy \& Pittsburgh Particle Physics, Astrophysics, and Cosmology Center (PITT PACC), University of Pittsburgh, 3941 O'Hara Street, Pittsburgh, PA 15260, USA}}
\newcommand{\Vtech}{\affiliation{Department of Physics, Virginia Tech, Blacksburg, VA 24061, USA}}
\newcommand{\IAC}{\affiliation{Instituto de Astrof{\'\i}sica de Canarias, E-38205 La Laguna, Tenerife, Spain}}
\newcommand{\Laguna}{\affiliation{Universidad de La Laguna, Dept. Astrof{\'\i}sica, E-38206 La Laguna, Tenerife, Spain}}
\newcommand{\UOak}{\affiliation{Homer L. Dodge Department of Physics and Astronomy, University of Oklahoma, 440 W. Brooks, Rm 100, Norman, OK 73019-2061, USA}}
\newcommand{\Hamburg}{\affiliation{Hamburger Sternwarte, Gojenbergsweg 112, D-21029 Hamburg, Germany}}
\newcommand{\PSI}{\affiliation{Planetary Science Institute, 1700 East Fort Lowell, Suite 106, Tucson, AZ 85719-2395 USA}}
\newcommand{\SETI}{\affiliation{SETI Institute, 339 Bernardo Ave, Suite 200, Mountain View, CA 94043, USA}}
\newcommand{\Hobart}{\affiliation{Physics Department, Hobart and William Smith Colleges, 300 Pulteney Street, Geneva, NY 14456, USA}}
\newcommand{\Cornell}{\affiliation{Department of Astronomy, Cornell University, 245 East Avenue, Ithaca, NY 14850, USA}}
\newcommand{\Athens}{\affiliation{IAASARS, National Observatory of Athens, Penteli 15236, Greece}}
\newcommand{\Turki}{\affiliation{Department of Physics and Astronomy, University of Turku, Vesilinnantie 5, 20500 Finland}}
\newcommand{\UMNAstro}{\affiliation{School of Physics and Astronomy, University of Minnesota, 116 Church Street S.E., Minneapolis, MN 55455, USA}}
\newcommand{\UVa}{\affiliation{Department of Astronomy, 530 McCormick Road, Charlottesville, VA 22904-4325, USA}}
\newcommand{\UTA}{\affiliation{Department of Physics, University of Texas at Arlington, Box 19059, Arlington, TX 76019, USA}}
\newcommand{\Konkoly}{\affiliation{Konkoly Observatory, CSFK, MTA Center of Excellence, Konkoly-Thege M. \'ut 15-17, Budapest, 1121, Hungary}}
\newcommand{\ELTE}{\affiliation{ELTE E\"otv\"os Lor\'and University, Institute of Physics and Astronomy, P\'azm\'any P\'eter s\'et\'any 1/A, Budapest, 1117 Hungary}}
\newcommand{\Szeged}{\affiliation{Department of Experimental Physics, University of Szeged, D\'om t\'er 9, Szeged, 6720, Hungary}}
\newcommand{\IAIFI}{\affiliation{The NSF AI Institute for Artificial Intelligence and Fundamental Interactions, USA}}
\newcommand{\UPadua}{\affiliation{Physics and Astronomy Department Galileo Galilei, University of Padova, Vicolo dell'Osservatorio 3, I-35122, Padova, Italy}}
\newcommand{\UWarwick}{\affiliation{Department of Physics, Gibbet Hill Road, University of Warwick, Coventry CV4 7AL, United Kingdom}}
\newcommand{\ING}{\affiliation{Isaac Newton Group of Telescopes, Apt. de Correos 368, E-38700 Santa Cruz de la Palma, Spain}}
\newcommand{\UNott}{\affiliation{School of Physics and Astronomy, University of Nottingham, University Park, Nottingham, NG7 2RD}}
\newcommand{\USurrey}{\affiliation{Mullard Space Science Laboratory, University College London, Holmbury St Mary, Dorking, Surrey RH5 6NT, United Kingdom}}
\newcommand{\USheffiled}{\affiliation{Department of Physics and Astronomy, University of Sheffield, Sheffield S3 7RH, UK}}
\newcommand{\Adler}{\affiliation{Adler Planetarium, 1300 S. DuSable Lake Shore Dr., Chicago, IL 60605, USA}}

%% file: authors.tex
\correspondingauthor{K. Azalee Bostroem}
\email{bostroem@arizona.edu}
\author[0000-0002-4924-444X]{K.\ Azalee Bostroem}
\Catalyst\UA
\email{bostroem@arizona.edu}
\author[0000-0001-8818-0795]{Stefano Valenti}
\UCD
\author[0000-0003-4102-380X]{David J.\ Sand}
\UA
\author[0000-0002-0744-0047]{Jeniveve Pearson}
\UA
\author[0000-0002-4022-1874]{Manisha Shrestha}
\UA
\author[0000-0003-0123-0062]{Jennifer E.\ Andrews}
\GeminiNorth
\author[0000-0003-0599-8407]{Luc Dessart}
\IAP
\author[0000-0002-3934-2644]{W.~V.~Jacobson-Gal\'{a}n}
\Hubble\Caltech
\author[0000-0002-9454-1742]{Brian Hsu}
\UA
\author[0000-0002-7352-7845]{Aravind P.\ Ravi}
\UCD
\author[0000-0002-1895-6639]{Moira Andrews}
\LCO\UCSB
\author[0000-0003-0528-202X]{Collin Christy}
\UA
\author[0000-0002-7937-6371]{Yize Dong \begin{CJK*}{UTF8}{gbsn}(董一泽)\end{CJK*}}
\UCD
\author[0000-0003-4537-3575]{Noah Franz}
\UA
\author[0000-0003-4914-5625]{Joseph Farah}
\LCO\UCSB
\author[0000-0003-3460-0103]{Alexei V. Filippenko}
\UCB
\author{Kiranjyot Gill}
\CfA
\author[0000-0003-2744-4755]{Emily T. Hoang}
\UCD
\author[0000-0002-0832-2974]{Griffin Hosseinzadeh}
\UCSD
\author[0000-0003-4253-656X]{D.\ Andrew Howell}
\LCO\UCSB
\author[0000-0003-0549-3281]{Daryl Janzen}
\USask
\author[0000-0001-5754-4007]{Jacob E.\ Jencson}
\IPAC
\author[0000-0001-8738-6011]{Saurabh W.\ Jha}
\Rutgers
\author[0000-0003-3108-1328]{Lindsey~A.~Kwok}
\CIERA
\author[0000-0001-9589-3793]{Michael Lundquist}
\Keck
\author[0009-0001-3106-0917]{Aidan Martas}
\UCB \UCD
\author[0000-0001-5807-7893]{Curtis McCully}
\LCO\UCSB
\author[0009-0008-9693-4348]{Darshana Mehta}
\UCD
\author[0000-0001-9570-0584]{Megan Newsome}
\UT
\author[0000-0003-0209-9246]{Estefania Padilla-Gonzalez}
\JHU
\author[0000-0002-7015-3446]{Nicolas E.\ Meza Retamal}
\UCD
\author[0000-0001-5510-2424]{Nathan Smith}
\UA
\author[0000-0001-8073-8731]{Bhagya M.\ Subrayan}
\UA
\author{Giacomo Terreran}
\Adler